\UseRawInputEncoding
\documentclass[lettersize,journal]{IEEEtran}
\usepackage{amsmath,amsfonts}
\usepackage{algorithmic}
\usepackage{algorithm}
\usepackage{array}
\usepackage[caption=false,font=normalsize,labelfont=sf,textfont=sf]{subfig}
\usepackage{textcomp}
\usepackage{stfloats}
\usepackage{url}
\usepackage{verbatim}
\usepackage{graphicx}
\usepackage{cite}
\usepackage{pifont}%
\usepackage{xcolor}
\usepackage{bm}

\makeatletter
\def\ps@IEEEtitlepagestyle{%
  \def\@oddfoot{\mycopyrightnotice}%
  \def\@evenfoot{}}
\def\mycopyrightnotice{%
  \hfill \footnotesize
  Accepted in: \textit{IEEE Internet of Things Journal
}, Oct. 14, 2025. 
  DOI: \texttt{10.1109/JIOT.2025.3623306}\hfill
}
\makeatother
\begin{document}

\title{Energy-Efficient Irregular RIS-aided UAV-Assisted Optimization:\\ A Deep Reinforcement Learning Approach}

\author{Mahmoud M. Salim, Khaled M. Rabie, \IEEEmembership{(Senior Member, IEEE)}, and Ali H. Muqaibel,
\IEEEmembership{(Senior Member, IEEE)}        
\thanks{All authors are with the Center for Communication Systems and Sensing, King Fahd University of Petroleum and Minerals, Dhahran 31261, Saudi Arabia. Also, Khaled M. Rabie is affiliated with the Computer Engineering Department, and Ali H. Muqaibel is with the Electrical Engineering Department at King Fahd University of Petroleum and Minerals, Dhahran 31261, Saudi Arabia. Mahmoud M. Salim is the corresponding author (email: mahmoud.elemam@kfupm.edu.sa).}}

\markboth{Journal of \LaTeX\ Class Files,~Vol.~14, No.~8, August~2021}%
{Shell \MakeLowercase{\textit{et al.}}: A Sample Article Using IEEEtran.cls for IEEE Journals}


\maketitle

\begin{abstract}
Reconfigurable intelligent surfaces (RISs) enhance unmanned aerial vehicles (UAV)-assisted communication by extending coverage, improving efficiency, and enabling adaptive beamforming. This paper investigates a multiple-input single-output system where a base station (BS) communicates with multiple single-antenna users through a UAV-assisted RIS, dynamically adapting to user mobility to maintain seamless connectivity. To extend UAV-RIS operational time, we propose a hybrid energy-harvesting resource allocation (HERA) strategy that leverages the irregular RIS ON/OFF capability while adapting to BS-RIS and RIS-user channels. The HERA strategy dynamically allocates resources by integrating non-linear radio frequency energy harvesting (EH) based on the time-switching (TS) approach and renewable energy as a complementary source. A non-convex mixed-integer nonlinear programming problem is formulated to maximize EH efficiency while satisfying quality-of-service, power, and energy constraints under channel state information and hardware impairments. The optimization jointly considers BS transmit power, RIS phase shifts, TS factor, and RIS element selection as decision variables. To solve this problem, we introduce the energy-efficient deep deterministic policy gradient (EE-DDPG) algorithm. This deep reinforcement learning (DRL)-based approach integrates action clipping and softmax-weighted Q-value estimation to mitigate estimation errors. Simulation results demonstrate that the proposed HERA method significantly improves EH efficiency, reaching up to 81.5\% and 73.2\% in single-user and multi-user scenarios, respectively, contributing to extended UAV operational time. Additionally, the proposed EE-DDPG model outperforms existing DRL algorithms while maintaining practical computational complexity.
\end{abstract}


\section{Introduction} \label{Sec_Intro}

\IEEEPARstart{T}{he} sixth generation (6G) wireless communication system has been driven by the increasing demand for seamless connectivity, high data rates, and low-latency services\cite{jiang-2021,wang-2023}. With the rapid proliferation of connected devices, including user equipment, Internet of Things (IoT) devices, and autonomous systems, future networks should address critical challenges such as spectrum limitations, energy efficiency (EE), and link reliability. Ensuring robust and adaptive network coverage is essential for various applications, including real-time data transmission, remote sensing, and intelligent automation.

Unmanned aerial vehicles (UAVs) have emerged as a transformative solution to address communication challenges, offering high mobility, flexible deployment, and rapid response capabilities \cite{Lin2025-Green}. They play a critical role in public safety scenarios, where conventional infrastructure is often unreliable or unavailable. By leveraging line-of-sight (LoS) communication and rapid deployment, UAVs can function as wireless relays, mobile base stations (BSs), or aerial network extenders, significantly improving connectivity in complex environments. However, UAV-assisted communications are highly susceptible to blockages from buildings and terrain, signal attenuation, and energy constraints, which limit their operational time and long-term effectiveness.

To enhance their communication reliability, UAVs have been integrated with reconfigurable intelligent surfaces (RISs), creating programmable radio environments by dynamically controlling signal reflections \cite{peng-2023}. Unlike conventional relays, passive RIS operates without amplification, improving spectral efficiency (SE) and EE without requiring additional radio frequency (RF) chains \cite{liu-2021S,kurma-2023-DRL}. UAV-mounted RISs provide full-space reflection and adaptive coverage, effectively mitigating signal blockages, extending communication range, and optimizing energy consumption.

Energy harvesting (EH) can play a vital role in extending the UAV-mounted RIS trip time by incorporating both RF and renewable energy (RE) sources\cite{peng-2023,kumar-2024,lee2020deep}. While RF EH provides a continuous but limited energy supply, RE sources generate power intermittently and can significantly contribute when available. By leveraging both energy sources, UAV-mounted RIS can enhance operational sustainability, reduce reliance on onboard batteries, and extend mission duration in energy-constrained environments.

Optimizing UAV-mounted RIS systems is challenging due to the highly coupled control variables, including UAV trajectory, RIS phase-shift design, and power allocation \cite{zhou-2023}. The dynamic nature of UAV movement and the need for real-time adaptation to channel variations add further complexity. Several optimization approaches have been explored. Model-based methods such as alternating optimization \cite{yu-2023}, successive convex approximation (SCA) \cite{long2020reflections}, and semidefinite relaxation \cite{xu-2022Semi} offer structured solutions but often involve high computational complexity. Heuristic techniques provide lower complexity alternatives but may compromise solution optimality \cite{dhuheir2024multi}. More recently, machine learning-based approaches, particularly deep reinforcement learning (DRL), have gained attention for handling complex, dynamic environments and large solution spaces efficiently \cite{samir-2021}. By leveraging DRL algorithms such as deep deterministic policy gradient (DDPG) and its variants, UAV-mounted RIS systems can dynamically adjust trajectory, beamforming, and RIS configurations. This enables improved SE, reduced power consumption, and enhanced overall network performance.

\subsection{Related Work} \label{Sec_Related_W}
 RIS improves wireless network performance by reconfiguring signal propagation to enhance energy and SE. In \cite{huang-2020}, the authors analyzed the joint design of transmit beamforming and RIS phase shifts in a multiple-input multiple-output (MIMO) multi-user (MU) system using DRL to optimize the sum-rate of the users. In \cite{kurma-2023-DRL}, the paper studied the SE-EE trade-off in a RIS-assisted full-duplex MU MIMO system using a DRL-based approach. The RIS phase shifts and power allocation were jointly optimized while considering partial channel state information (CSI), co-channel interference, and residual self-interference to enhance system performance. The work in \cite{cao-2023-DRL} examined a RIS-assisted multiple-input single-output (MISO) symbiotic radio system with finite blocklength transmissions. An EE maximization problem was formulated by jointly optimizing the BS beamforming vector and RIS phase shifts, and a DRL-based framework was proposed to solve the non-convex problem. Additionally, a distributed RIS architecture was explored to improve EE by coordinating the joint optimization of transmit power and RIS scheduling using SCA \cite{yang-2021}.
 
The integration of RIS with EH enhances EE in wireless networks by optimizing EH and management. Studies focus on different aspects including power control, phase shift optimization, and system efficiency. The work in \cite{chen-2024} optimized an active RIS-assisted downlink system with time-switching (TS) EH. It jointly configured RIS phase shifts, time allocation, and power control using a two-step convex optimization approach to enhance EE. In \cite{mohammadi-2024}, the authors investigated a system of RIS-assisted massive MIMO enabling wireless power transfer to energy-limited IoT devices. The BS transmit power and RIS reflection coefficients were optimized utilizing SCA, considering CSI errors to enhance harvested energy. In \cite{ghose-2024}, the authors studied a passive RIS-assisted device-to-device cognitive radio network, where a batteryless secondary transmitter harvested energy from RF signals. They maximized EE by jointly optimizing transmit power and RIS beamforming via an alternating optimization algorithm. The work in \cite{kumar-2024} analyzed a wireless-powered IoT network assisted by passive and active RIS, considering a nonlinear EH model. They derived analytical expressions for outage probability, sum throughput, and EE. In \cite{sharma-2024}, the authors optimized power splitting and transmit power in RIS-assisted simultaneous wireless information and power transfer (SWIPT) systems. This came with a phase-dependent amplitude model for RIS reflectivity employing fractional programming-based solutions to maximize data rate and EH. The RIS-assisted SWIPT networks were investigated in \cite{zhang-2023-RSMA}, formulating an EE maximization problem considering quality-of-service (QoS) constraints. A proximal policy optimization (PPO)-based DRL approach was proposed to jointly optimize beamforming, power splitting, and discrete phase shifts. In \cite{lee2020deep}, a dynamic adaptable EH model was integrated into the RIS, allowing it to harvest wireless energy from various sources and improve EE in RIS-assisted cellular networks. Expanding on RIS-equipped EH, the study in \cite{wu-2020} showcased that a RIS-enabled approach is able to effectively reduce the access point’s transmission power by strategically configuring passive phase shifts and optimizing transmitter precoding.

The combination of RIS with UAVs has garnered significant interest for enhancing energy-efficient and resilient wireless communications. Various studies have explored different optimization strategies to improve UAV operational time, trajectory control, and service scheduling in RIS-assisted UAV networks. In \cite{Lin2025-Green}, the authors studied EE optimization in RIS-assisted fixed-wing UAV communications by jointly optimizing UAV trajectory and service scheduling. They utilized a hybrid DRL approach to enhance exploration efficiency and action space accuracy. An active RIS-assisted UAV communication system to enhance uplink connectivity in emergency scenarios was proposed in \cite{hu-2024UAV}. DRL was used to jointly optimize UAV trajectory and active RIS phase control for EE maximization. The UAV-enabled RIS paradigm in \cite{samir-2021} successfully mitigated the interruption between the IoT devices and the BS; however, the battery-operated UAV posed a difficulty due to its restricted operational time. An iterative SCA approach was introduced to boost the secure EE of UAV-enabled RIS networks by jointly adjusting the phase shifts of reflective elements, managing transmit power, and optimizing UAV trajectory \cite{long2020reflections}. In \cite{9771999}, a dual-domain based on a conventional TS EH scheme utilizing DDPG was introduced to improve the endurance of UAV-enabled RIS networks. However, this approach was evaluated only for a single UT and faced reinforcement learning underestimation issues, leading to suboptimal EH efficiency.

\subsection{Motivations and Contributions} \label{Sec_Motiv.}
Extending the operational time of UAV-RIS-assisted networks is a critical challenge, as UAVs are inherently constrained by limited onboard energy. Maximizing EH efficiency by increasing harvested energy and reducing reliance on conventional power sources is essential to overcoming this limitation. Table \ref{Tab:Contr} provides a comparative analysis of the most recent works on RIS-assisted communication networks, highlighting the advancements and methodologies utilized in the literature. Notably, "Deployment" indicates whether the UAV and user equipment are fixed or mobile, while "RIS + UAV" denotes whether the RIS is integrated with a UAV or deployed in a fixed location. "Irregular RIS" specifies whether RIS elements can dynamically switch on/off for EH or signal reflection. 
\begin{table*}[t]
\caption{Comparison with Existing Works in the Literature}
\centering
\resizebox{\textwidth}{!}{
\begin{tabular}{ccccccccccc}
\hline
\textbf{Ref.}               & \textbf{Deployment} & \textbf{MUs} & \textbf{RIS Type} & \textbf{UAV+RIS} & \textbf{EH}   & \textbf{Irregular RIS} & \textbf{iCSI} & \textbf{HI} & \textbf{Optimization Scheme} & \textbf{Metric(s)} \\ \hline
\cite{huang-2020}          & Fixed               & $\checkmark$ & Passive           & $\times$                                                      & $\times$      & $\times$                                                      & $\times$      & $\times$    & DDPG                         & Sum-rate           \\ \hline
\cite{kurma-2023-DRL}      & Fixed               & $\checkmark$ & Passive           & $\times$                                                      & $\times$      & $\times$                                                      & $\checkmark$  & $\times$    & DDPG                         & SE \& EE           \\ \hline
\cite{cao-2023-DRL}        & Fixed               & $\checkmark$ & Passive           & $\times$                                                      & $\times$      & $\times$                                                      & $\times$      & $\times$    & Double DDPG                  & EE                 \\ \hline
\cite{yang-2021}           & Mobile              & $\checkmark$ & Active            & $\times$                                                      & $\times$      & $\times$                                                      & $\times$      & $\times$    & SCA                          & EE                 \\ \hline
\cite{chen-2024}           & Fixed               & $\times$     & Active            & $\times$                                                      & Linear TS     & $\times$                                                      & $\times$      & $\times$    & Convex optimization          & EE                 \\ \hline
\cite{mohammadi-2024}      & Fixed               & $\times$     & Passive           & $\times$                                                      & Non-linear TS & $\times$                                                      & $\checkmark$  & $\times$    & SCA                          & Data rate \& EH    \\ \hline
\cite{kumar-2024}          & Fixed               & $\checkmark$ & Active/Passive    & $\times$                                                      & Non-linear TS & $\times$                                                      & $\times$      & $\times$    & Closed-form expressions      & OP, SE, \& EE      \\ \hline
\cite{sharma-2024}         & Fixed               & $\checkmark$ & Passive           & $\times$                                                      & Non-linear PS & $\times$                                                      & $\times$      & $\times$    & Fractional programming       & Data rate \& EH    \\ \hline
\cite{zhang-2023-RSMA}     & Fixed               & $\checkmark$ & Passive           & $\times$                                                      & Non-linear PS & $\times$                                                      & $\times$      & $\times$    & PPO-based DRL                & EE                 \\ \hline
\cite{9771999}             & Fixed               & $\checkmark$ & Passive           & $\checkmark$                                                  & Non-linear TS & $\checkmark$                                                  & $\times$      & $\times$    & DDPG                         & EH Efficiency      \\ \hline
\cite{lee2020deep}         & Fixed               & $\checkmark$ & Passive           & $\times$                                                      & Generic       & $\checkmark$                                                  & $\times$      & $\times$    & DRL                          & EE                 \\ \hline

\cite{Lin2025-Green}       & Fixed               & $\checkmark$ & Passive           & $\checkmark$                                                  & $\times$      & $\times$                                                      & $\times$      & $\times$    & Hybrid DRL                   & EE                 \\ \hline
\cite{hu-2024UAV}          & Fixed               & $\checkmark$ & Active            & $\checkmark$                                                  & $\times$      & $\times$                                                      & $\times$      & $\times$    & PPO-based DRL                & SE \& EE           \\ \hline
\cite{samir-2021}          & Mobile              & $\checkmark$ & Passive           & $\checkmark$                                                  & $\times$      & $\times$                                                      & $\times$      & $\times$    & PPO-based DRL                & Age-of-Information \\ \hline
\cite{long2020reflections} & Mobile              & $\checkmark$ & Passive           & $\checkmark$                                                  & $\times$      & $\times$                                                      & $\times$      & $\times$    & SCA                          & Secure EE          \\ 
\hline
\textbf{Proposed} & Mobile              & $\checkmark$ & Passive           & $\checkmark$                                                  & Non-linear TS RF + RE     & $\checkmark$                                                      & $\checkmark$      & $\checkmark$    & Improved Dual DDPG                          & EH efficiency          \\ 
\hline
\end{tabular}
\label{Tab:Contr}
}
\end{table*}

As shown in the table, previous research has primarily focused on maximizing system SE or EE while minimizing energy consumption in RIS-assisted networks. While some studies have ensured QoS requirements for users, they have yet to thoroughly explore UAV operational time maximization represented by EH efficiency. Moreover, most existing works rely solely on RF signals for EH, which proves insufficient even for short UAV trips. A more sustainable approach integrates RE sources, such as solar cells, with RF EH technology, providing a practical and reliable energy supply for UAV operations. Additionally, leveraging the irregular RIS ON/OFF capability in combination with TS EH, where RIS elements dynamically switch between EH and signal reflection, has not been sufficiently explored. Despite significant advancements in UAV-RIS systems, most existing studies have overlooked critical impairments, including CSI imperfections, RF EH non-linearity, and hardware impairments (HIs), which significantly impact system performance and efficiency.

Therefore, motivated by the above, we present a practical MISO system of a BS communicating with MUs, each with a single antenna, through a hybrid RF and RE harvesting UAV-RIS unit. This model leverages the irregular RIS ON/OFF capability in combination with TS EH, allowing RIS elements to intelligently switch between EH and signal reflection. By considering practical constraints such as RF EH non-linearity, CSI imperfections, and HIs, our approach ensures a more efficient and sustainable UAV-RIS-assisted communication network. To the best of our knowledge, this is the first practical approach to combine all the aforementioned considerations for enhancing the EH efficiency of UAV-assisted RIS communication.

Specifically, the main contributions of this paper can be summarized as follows:
\begin{itemize}
\item The hybrid energy-harvesting resource allocation (HERA) strategy is proposed to exploit the irregular RIS ON/OFF capability in combination with non-linear TS RF and RE harvesting, prolonging the UAV-RIS unit's operational time. The proposed strategy accounts for the different nature of BS-RIS and RIS-MUs channel models.
\item A non-convex mixed-integer nonlinear programming (MINLP) problem is formulated to maximize the EH efficiency subject to the MUs' QoS and EH constraints with impairments. This jointly optimizes the BS transmit power, RIS phase shift, TS parameter, and dynamic selection of reflective elements.
\item To tackle the formulated problem, the energy-efficient DDPG (EE-DDPG) algorithm is proposed as an enhanced variant of the twin delayed DDPG (TD3) algorithm introduced in \cite{fujimoto2018}. The EE-DDPG algorithm incorporates advanced features such as action clipping and the softmax function to mitigate underestimation and overestimation issues while dynamically adapting to user mobility for optimal resource allocation.
\item Simulation results validate the effectiveness of the HERA strategy in extending the EH efficiency of UAV-RIS systems, while the proposed EE-DDPG outperforms other DRL benchmarks.
\end{itemize}
\subsection{Paper Organization}
The remainder of this paper is structured as follows. Section \ref{Sec_SM} presents the system model, detailing the UAV-RIS architecture, EH framework, and transmission model. Section \ref{Sec_PF} formulates the optimization problem, defining key constraints and decision variables for maximizing EH efficiency while ensuring communication reliability. In Section \ref{Sec_Proposed_Algorithm}, the proposed EE-DDPG algorithm is introduced as a DRL-based solution to deal with the formulated problem. Section \ref{Sec_Simu} evaluates the proposal performance through numerical simulations, comparing it with existing benchmarks. Finally, Section \ref{Sec_Conclusion} concludes the paper and outlines potential future research directions.
\section{System Model} \label{Sec_SM}
\begin{figure}[!th]
\centering
\includegraphics[width=0.9\columnwidth]{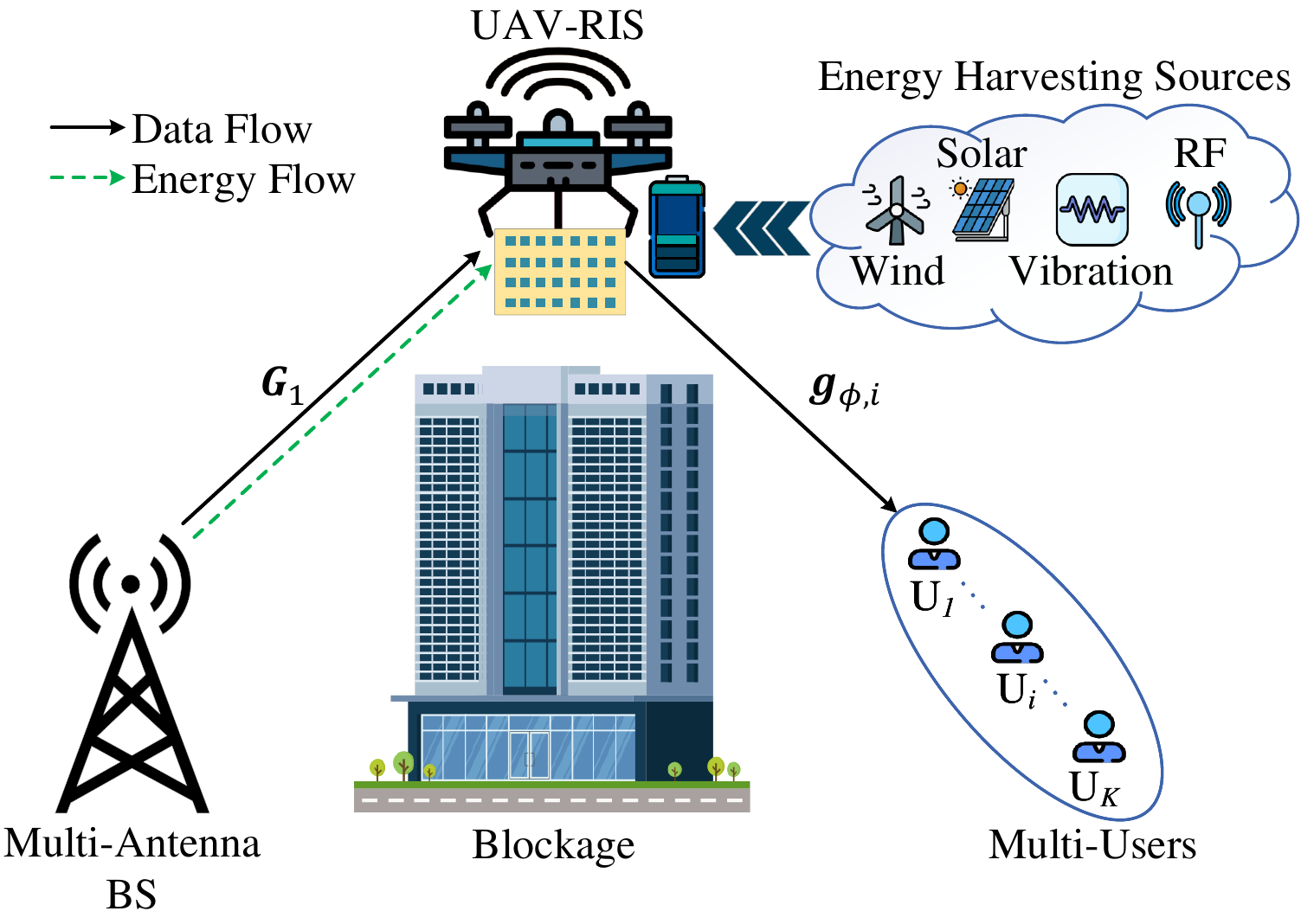}
\caption{Proposed system model of MU communication with a serving multiple-antenna BS via a UAV-RIS unit in the absence of LoS component.}
\label{Fig_SM}
\end{figure}

Fig. \ref{Fig_SM} illustrates a MISO system consisting of a BS, a hybrid RF and RE harvesting UAV-RIS unit, and \( K \) single-antenna users, represented by the set \( \mathbb{K} = \{1, 2, \dots, K\} \). The RIS comprises \( \mathit{L} = \mathit{M} \times \mathit{N} \) reflective meta-surfaces. To sustain its operation, the UAV-RIS harvests energy from RF signals received from the BS as well as RE sources such as solar, wind, and vibration. Direct communication between the BS and users is obstructed due to LoS blockage, which is common in public safety scenarios. To address this, the UAV functions as a mobile BS, while the RIS dynamically adjusts the phase shifts of its meta-surfaces via an integrated microcontroller to enhance signal propagation.

The total communication period is partitioned into \( T \) equal time slots, denoted by \( \mathbb{T} = \{1, 2, \dots, t, \dots, T\} \). Each time slot \( t \) consists of two phases, which are the EH phase of duration \( \tau(t) \) and the information transmission phase lasting \( (1-\tau(t)) \). The position of user \( k \) at time slot \( t \) is given by \( \mathcal{P}^k(t) = (x^k(t), y^k(t), z^k(t)) \), where \( z^k(t) \) represents the user's antenna height, and \( (x^k(t), y^k(t)) \) denotes its horizontal location in a Cartesian coordinate system with the BS at the origin. The BS, equipped with \( \mathit{A} \) antennas, transmits signals to the UAV-RIS, assuming that users receive signals solely via UAV-RIS reflection. Each RIS element, located at row \( i \) and column \( j \), is denoted as \( \Phi_{i,j} \) and positioned at \( \mathcal{P}^\phi_{i,j}(t) = (x^\phi_{i,j}(t), y^\phi_{i,j}(t), z^\phi_{i,j}(t)) \). The UAV's trajectory inherently determines the positioning of these elements. 

To account for practical constraints, we incorporate CSI imperfections following \cite{bisen-2021}. The minimum mean square error channel estimator models the estimated channel gain as \( g = h + e \), where \( h \sim \mathcal{CN}(0, \sigma^2) \) represents the actual channel gain with variance \( \sigma^2 \), and \( e \sim \mathcal{CN}(0, \zeta^2) \) denotes the estimation error with variance \( \zeta^2 \). Furthermore, considering hardware impairment (HI) as in \cite{sesia-2011,arzykulov-2019}, the distortion noise between any transmitter \( a \) and receiver \( b \) is modeled as \( \Psi_{a-b} \sim \mathcal{CN}(0, \psi^2_{a-b}) \), where \( \psi_{a-b} = \sqrt{\psi_{a} + \psi_{b}} \).  The most important symbols used throughout this paper are summarized in Table \mbox{\ref{Tab:Notation}}.
\begin{table}[h]
\caption{List of Symbols}
\label{Tab:Notation}
\centering
\setlength{\tabcolsep}{0.75pt}
\renewcommand{\arraystretch}{1.0}
\scalebox{0.8}{
\begin{tabular}{|c|c|}
\hline
\textbf{Symbol}& 
\textbf{Description}\\

\hline
$\mathit{L}=\mathit{M} \times \mathit{N}$& 
Number of RIS elements\\
\hline
$\mathbb{K}$& 
Set of users\\
\hline
$K$& 
Number of users\\
\hline
$t$& 
time slot\\
\hline
$\mathbb{T}$& 
Set of time slots\\
\hline
$T$& 
Number of time slots\\
\hline
$\mathit{A}$& 
Number of BS antennas\\

\hline
$\mathcal{P}$& 
Node position\\
\hline
$\Phi$& 
RIS element\\
\hline
$h$& 
Channel gain with variance $\sigma^2$\\
\hline
$e$& 
Channel estimation error with variance $\zeta^2$\\
\hline
$g$& 
Estimated channel gain\\

\hline
$\Psi_{ab}$& 
Combined HI distortion noise between $a$ and $b$\\
\hline
$\varepsilon^{RE}$& 
Instantaneous RF energy amount at UAV-RIS\\
\hline
$\lambda$& 
RE rate of arrival\\
\hline
$\tau$& 
EH phase duration\\

\hline
$\mathbf{S}$& 
BS transmitted signal\\

\hline
$\mathbf{V}_k$& 
Precoding vector of user $k$\\
\hline
$\mathcal{C}_k$& 
Transmitted signal of user $k$\\

\hline
$p_k$& 
Power allocated to user $k$\\
\hline
$\varepsilon^{RF-NL}$& 
Instantaneous non-linear RF energy amount UAV-RIS\\
\hline
$P_{sat}$& 
Saturation harvesting power of RF
energy harvester\\

\hline
$P^{RF}$& 
Total received RF power at the UAV-RIS \\

\hline
$\mathbf{g}_{i,j}$& 
channel vector between BS antennas and RIS element $\Phi_{i,j}$\\

\hline
$\mathbf{G}_{1}$& 
Small-scale fading matrix\\

\hline
$P_{BS}$& 
BS transmit power\\

\hline
$\mathcal{PL}$& 
Path-loss\\

\hline
$\mathcal{P}r(x)$& 
Probability of $x$\\
\hline
$\varphi$& 
Non-LoS attenuation \\

\hline
$C_X \text{ and } C_Y$& 
Environment-dependent constants\\

\hline
$\theta_{i,j}$& 
Elevation angle between BS and RIS element $\Phi_{i,j}$\\

\hline
$\varepsilon^h$& 
Total energy harvested by UAV-RIS unit\\
\hline
$\varepsilon^{r}$& 
Residual energy of UAV-RIS unit\\
\hline
$\varepsilon^{c}$& 
Consumed energy by UAV-RIS unit\\
\hline

$\varepsilon^{h}_{HERA}$& 
HERA strategy harvested energy\\
\hline
$ \beta^k_{i,j}$& 
Resource allocation binary variable\\
\hline
$\mathbf{g}_{\phi,k}$& 
Channel vector between UAV-RIS unit and user $k$\\

\hline
$\bm{\theta}$& 
Diagonal reflection coefficient matrix\\

\hline
$\rho$& 
Reflection amplitude coefficient\\

\hline
$Y_k$& 
Received RF signal at user $k$\\

\hline
$\sigma_k$& 
Additive white
Gaussian noise at user $k$\\

\hline
$\hat{\mathbf{G}}_{r,k}$& 
Channel matrix between UAV-RIS unit and user $k$\\

\hline
$\upsilon$& 
Reference path loss\\
\hline
$\alpha$& 
Path loss exponent for BS-RIS links\\
\hline
$\nu$& 
Path loss exponent for RIS-user links\\

\hline
$\mathbf{g}_{\phi,k}^{\text{LoS}}$& 
Deterministic LoS component\\

\hline
$\mathbf{g}_{\phi,k}^{\text{NLoS}}$& 
Rayleigh fading for the non-NLoS component\\

\hline
$\Gamma_k$& 
SNR at user $k$\\

\hline
$R_k$& 
Data rate at user $k$ (bits/s/Hz)\\
\hline
\end{tabular}}
\label{Tab:Notation}
\end{table}
\subsection{Energy Harvesting Model} \label{Sec_EH}
Here, we first introduce the conventional TS strategy, where all RIS elements are either fully dedicated to EH or signal reflection within each time slot. Building on this foundation, we propose the HERA strategy, which dynamically reconfigures RIS elements to achieve a balanced trade-off between EH and signal reflection. This adaptive approach enhances EE and ensures sustained UAV-RIS operation while optimizing signal transmission.
\subsubsection{Conventional Non-Linear Time-Switching Strategy} \label{Sec_ConvTS}

We consider a versatile EH scheme to enhance the EH efficiency of the UAV-RIS unit and prolong its battery operational time. Specifically, we adopt an advanced RF/RE EH model, where the UAV-RIS is equipped with dual EH mechanisms to collect energy from both RF signals and RE sources, ensuring continuous and sustainable operation. For RE harvesting, we assume that the UAV-RIS unit gathers RE at the start of each time slot \( t \), modeled using a discrete-time EH framework \cite{salim-2020, salim-2024}. This framework follows an independent uniformly distributed process with energy arrivals \( \varepsilon^{RE} \in [0, \mu] \) Joules. Additionally, the arrival times of the harvested energy follow a Poisson distribution with an average rate of \( \lambda \) J/s \cite{salim-2019Op,salim-2023AS}.

Alongside RE harvesting, the RIS is equipped with RF EH capability, enabling its elements to collect and utilize harvested energy for operation. Following the TS protocol, each time slot \( t \) is divided into two phases: the EH phase \( \tau(t) \) and the information transmission phase \( 1-\tau(t) \). During the EH phase, all reflective elements are dedicated to EH. Upon completion, the system seamlessly transitions into the information transmission phase, where the meta-surfaces reflect the incoming signals. As per \cite{peng-2023}, the BS transmits the signal as  
\begin{equation}
\mathbf{S} = \sum_{k \in \mathbb{K}} \mathbf{V}_k \mathcal{C}_k,
\label{eq:BS_trans_Sig}
\end{equation}
where \( \mathbf{V}_k \in \mathbb{C}^{D \times 1} \) is the precoding vector, and \( \mathcal{C}_k \) represents the transmitted signal for the \( k \)-th user, with \( \mathcal{C}_k \sim \mathcal{CN}(0,1) \). Consequently, the total transmission power at the BS satisfies  
\begin{equation}
\mathbb{E} \left[ \mathbf{S}^H \mathbf{S} \right] = \sum_{k \in \mathbb{K}} \|\mathbf{V}_k\|^2,
\label{eq:BS_trans_Pw}
\end{equation}
where \( \|\cdot\| \) denotes the Euclidean norm, and the power allocated to user \( k \) is given by \( p_k = \|\mathbf{V}_k\|^2 \). For practical considerations, the RF harvested energy follows a nonlinear EH model \cite{sharma-2022,Semiha2023}, expressed as  
\begin{equation}
    \varepsilon^{\text{RF-NL}} = \frac{\Omega - P_{sat} \Delta}{1 - \Delta},
\label{eq:P_RF-NL}    
\end{equation}
where \( P_{sat} \) represents the saturation power of the energy harvester, {\footnotesize \( \Omega = P_{sat}/(1+\text{exp}(-c(P^{RF}-d))) \)}, and {\footnotesize \( \Delta=1/(1+\text{exp}(cd)) \)}. The constants \( c \) and \( d \) are related to hardware characteristics. The total received RF power at the UAV-RIS at time slot \( t \) is given by  
\begin{equation}
P^{RF}(t) = \tau (t) \sum\limits_{i=1}^{\mathit{M}} \sum\limits_{j=1}^{\mathit{N}} \| \mathbf{g}_{i,j}^H\mathbf{S} \|^2,
\end{equation}
where, {\footnotesize \( \mathbf{g}_{i,j} = \left( g_{i,j}^{(1)}, \dots, g_{i,j}^{(\mathfrak{a})}, \dots, g_{i,j}^{(\mathit{A})} \right) \)} represents the channel vector between the BS antennas \( \mathit{A} \) and RIS element \( \Phi_{i,j} \), following the air-to-ground path loss model \cite{peng-2023,peng-2021-Leo}. The small-scale fading in $\mathbf{G_1} = \left( \mathbf{g}_{1,1}^H, \dots, \mathbf{g}_{1,\mathit{N}}^H, \dots, \mathbf{g}_{\mathit{M},\mathit{N}}^H \right) \in \mathbb{C}^{\mathit{A} \times \mathit{L}}$ follows Rayleigh fading. The BS’s total transmission power is given by $P_{BS} = \mathbb{E} \left[ \mathbf{S}^H \mathbf{S} \right]$. The path loss \( \mathcal{PL}_{i,j} \) between the BS and RIS element \( \Phi_{i,j} \) is modeled as \cite{mei-2021}  
\begin{align}
\mathcal{PL}_{i,j} &= \left( \mathcal{P}r_{i,j}(\text{LoS}) + (1 - \mathcal{P}r_{i,j}(\text{LoS})) \varphi \right) \nonumber \\
& \times  \left( \sqrt{|x_{i,j}(t)|^2 + |y_{i,j}(t)|^2 + |z_{i,j}(t)|^2} \right)^{-\alpha},
\end{align}
where \( \alpha \) denotes the path loss exponent, \( \varphi \) is the additional attenuation due to non-LoS propagation, and \( \mathcal{P}r_{i,j}(\text{LoS}) \) is the probability of a LoS connection between the BS and RIS element \( \Phi_{i,j} \). According to \cite{peng-2023,lei-2021}, this probability is given by  
\begin{equation}
\mathcal{P}r_{i,j}(\text{LoS}) = \frac{1}{1 + C_X \exp{(-C_Y (\theta_{i,j} - A))}},
\end{equation}
where \( C_X \) and \( C_Y \) are environment-dependent constants \cite{al2014modeling}. The elevation angle \( \theta_{i,j} \) between the BS and RIS element \( \Phi_{i,j} \) is expressed as  
\begin{equation}
\theta_{i,j} = \frac{180}{\pi} \sin^{-1} \left( \frac{z_{i,j}(t)}{\sqrt{|x_{i,j}(t)|^2 + |y_{i,j}(t)|^2 + |z_{i,j}(t)|^2}} \right).
\end{equation}

\subsubsection{Hybrid Energy-Harvesting Resource Allocation Strategy} \label{Sec_HERA}
To further enhance UAV operational efficiency and extend its lifetime, we build upon the conventional TS strategy by introducing the HERA strategy \cite{fadel2024irregular,sohail2024irregular}. Unlike conventional TS models where RIS elements serve fixed roles, the proposed approach dynamically reconfigures the RIS elements, leveraging their irregular structure for efficient EH and signal reflection.

\begin{figure}[!t]
\centering
\includegraphics[width=0.9\columnwidth]{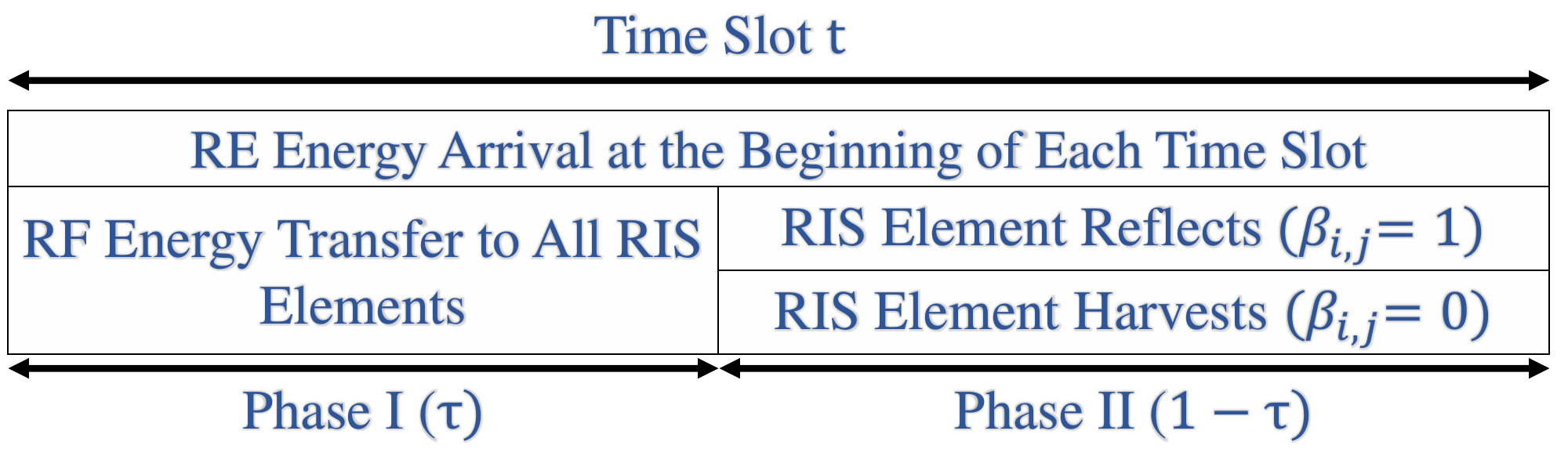}
\caption{Proposed HERA strategy for the UAV-RIS unit.}
\label{Fig_EHM}
\end{figure}

Fig. \ref{Fig_EHM} illustrates the operation of the HERA strategy within each time slot, which consists of two phases. During phase I, all RIS elements are dedicated to EH. In phase II, a subset of RIS elements continues EH, while the remaining elements are dynamically assigned for signal reflection. This hybrid allocation ensures an additional energy supply for the UAV-RIS unit while simultaneously optimizing signal transmission. As a secondary power source, RE arrives at the beginning of each time slot.
At each time slot \( t \), the total harvested energy under the HERA strategy is given by  
\begin{align}
    \varepsilon^h_{HERA}(t) &= \tau(t) \varepsilon^{\text{RF-NL}}(t) \nonumber \\
    & + (1 - \tau(t)) \sum_{i=1}^{\mathit{M}} \sum_{j=1}^{\mathit{N}} \left( 1 - \sum_{k \in \mathbb{K}} \beta^k_{i,j} \right) \varepsilon^{\text{RF-NL}}(t) \nonumber \\
    & + \varepsilon^{\text{RE}}(t),
    \label{eq:HERA_EH}
\end{align}
where {\footnotesize\( \beta_{i,j}^k \in \{0, 1\} \)} and {\footnotesize\( \sum_{k \in \mathbb{K}} \beta_{i,j}^k \leq 1 \)}, for all {\footnotesize\( i \in [1, \mathit{M}], j \in [1, \mathit{N}], k \in \mathbb{K} \)}. Specifically, \(\beta_{(i,j)}^k = 1\) indicates that the RIS element \(\Phi_{i,j}\) is assigned for reflection to user \( k \), whereas \(\beta_{(i,j)}^k = 0\) means it remains in EH mode. This adaptive element selection strategy enables the UAV-RIS to balance EH and signal reflection, thereby optimizing both EE and communication reliability.

The proposed model follows a harvest-transmit-use strategy \cite{salim-2022}, where harvested energy is stored for future utilization. To ensure operational feasibility, we impose the energy causality constraint, which ensures energy is used only after it has been harvested, and the battery overflow constraint, which respects the UAV-RIS unit’s maximum energy storage capacity \( \varepsilon_{\max} \). Consequently, the harvested and residual energy of the UAV-RIS at time slot \( t \) are given as 
\begin{equation}
    \varepsilon^{h}(t) = \varepsilon^{\text{RF-NL}}(t) + \varepsilon^{\text{RE}}(t),
\end{equation}

\begin{equation}
    \varepsilon^{r}(t) = \min \left\{ \varepsilon_{\max}, \varepsilon^{r}(t-1) +  \varepsilon^{h}(t) -  \varepsilon^{c}(t) \right\},
\end{equation}
where \( \varepsilon^{c}(t) \) represents the energy consumed by the UAV-RIS for operation and transmission at time slot \( t \). 
\subsection{Transmission Model} \label{Sec_Trans_Model}
In this work, we consider a UAV-RIS system that performs passive reflective beamforming to enhance communication efficiency. During the information transmission phase of the $t$-th time slot, the channel vector {\footnotesize$\mathbf{g}_{\phi,k} = \big(g_{1,1}(k), \dots, g_{1,\mathit{N}}(k), \dots, g_{\mathit{M},\mathit{N}}(k) \big)$} is the equivalent baseband channels from the UAV-RIS to the $k$-th user. Similarly, {\footnotesize$\mathbf{G_1} \in \mathbb{C}^{\mathit{A} \times \mathit{L}}$} denotes the channel matrix from the BS to the UAV-RIS. The UAV-RIS dynamically modifies the phase shifts of its elements to passively redirect the incoming signals.

To model this behavior, we define a diagonal reflection coefficient matrix $\bm{\theta}$, capturing the phase shift and amplitude reflection of each meta-surface element as $ \bm{\theta} = \text{diag}(\rho_1 e^{j\theta_1}, \dots, \rho_\mathit{L} e^{j\theta_\mathit{L}}) \in \mathbb{C}^{\mathit{L} \times \mathit{L}}$. Here, $j = \sqrt{-1}$ represents the imaginary component, $\theta_l \in (0, 2\pi)$ is the phase shift of the $l$-th reflective element, and $\rho_l \in [0,1]$ is its reflection amplitude coefficient. For maximum reflection efficiency, we assume $\rho_l = 1$ for all $l$, enabling independent phase control of each RIS element. 

Considering UAV-RIS-assisted communication, the received RF signal at user $k$ via the BS-RIS-user $k$ link is given by  
\begin{equation}
    Y_k = \hat{\mathbf{G}}_{\phi,k}^{H} \bm{\theta}^{H} \mathbf{G_1}^H (\mathbf{S} + \bm{\Psi_{BS-U_k}}) + n_k, \quad k \in \mathbb{K},
\end{equation}
where \( \bm{\Psi_{BS-U_k}} \sim \mathcal{CN}(0, \bm{\psi_{BS-U_k}}^2) \) represents the cumulative HI noise affecting the entire transmission path from the BS to user $k$, and $n_k \sim \mathcal{CN}(0, \sigma_k^2)$ denotes the additive white Gaussian noise at user $k$. The matrix $\hat{\mathbf{G}}_{\phi,k}$ represents the channel from the UAV-RIS to user $k$, incorporating RIS reflection assignment, and is given by  
\begin{equation}
\hat{\mathbf{G}}_{r,k} =
\begin{bmatrix}
\beta_{1,1}^k g_{1,1}(k) & \dots & \beta_{1,\mathit{N}}^k g_{1,\mathit{N}}(k) \\
\vdots & \ddots & \vdots \\
\beta_{\mathit{M},1}^k g_{\mathit{M},1}(k) & \dots & \beta_{\mathit{M},\mathit{N}}^k g_{\mathit{M},\mathit{N}}(k)
\label{eq:RA_matrix}
\end{bmatrix}.
\end{equation}

The channel $\mathbf{g}_{\phi,k}$ accounts for both path loss and small-scale fading. The path loss between the UAV-RIS and the users is modeled as  
    $\upsilon \left(\frac{d_{i,j}^k (t)}{d_{ref}} \right)^{-\nu}$,
where $\nu$ denotes the path loss exponent for the RIS-user links, $d_{i,j}^k (t) = \|\mathcal{P}_k (t) - \mathcal{P}^\phi_{i,j} (t)\|_2$ is the geometric distance from the reflective element $\Phi_{i,j}$ to user $k$, and $\upsilon$ is the reference path loss at $d_{ref} = 1$ m. For small-scale fading, we assume a Rician fading model with a Rician factor of $K_{\text{rician}} = 10$, given by 
$\mathbf{g}_{\phi,k} = \sqrt{\frac{K_{\text{rician}}}{1+K_{\text{rician}}}} \mathbf{g}_{\phi,k}^{\text{LoS}} + \sqrt{\frac{1}{1+K_{\text{rician}}}} \mathbf{g}_{\phi,k}^{\text{NLoS}}$,
where $\mathbf{g}_{\phi,k}^{\text{LoS}}$ represents the deterministic LoS component, while $\mathbf{g}_{\phi,k}^{\text{NLoS}}$ follows Rayleigh fading for the non-LoS component.

Following the interference cancellation assumption in \cite{tang2020joint}, each user can perfectly eliminate undesired signals from other RIS-user $k$ links before decoding its own signal. The resulting signal-to-noise ratio (SNR) at user $k$ is defined as  
\begin{equation} 
\Gamma_k = \frac{\big| \hat{\mathbf{G}}_{\phi,k}^{H} \bm{\theta}^{H} \mathbf{G_1}^H \mathcal{C}_k \big|^2} {\big| \hat{\mathbf{G}}_{\phi,k}^{H} \bm{\theta}^{H} \mathbf{G_1}^H \psi_{BS-U_k} \big|^2 + \sigma_k^2 }. 
\label{eq:SNR_user_k} 
\end{equation}

Using Shannon’s capacity formula, the achievable data rate (bits/s/Hz) at user $k$ during  $t$-th time slot is  
\begin{equation}
R_k(t) = (1 - \tau (t)) B \log_2 (1 + \Gamma_k), \quad k \in \mathbb{K}, t \in \mathbb{T},
\end{equation}
where $B$ represents the system bandwidth. The overall EH efficiency of the UAV-RIS unit is then given by  
\begin{align}
    \varepsilon \varepsilon &= \frac{\varepsilon^h_{HERA}(t)}{\varepsilon^{Rx}_{UAV-RIS}(t)},
\end{align}
where $\varepsilon^{Rx}_{UAV-RIS}(t) = \sum\limits_{i=1}^{\mathit{M}} \sum\limits_{j=1}^{\mathit{N}} \left\| \mathbf{g}_{i,j}^H \mathbf{S} \right\|^2$
represents the total received energy from incident RF signals at each time slot \( t \).  
\section{Problem Formulation} \label{Sec_PF}
We formulate an optimization problem of a MISO system, where a BS with \( \mathit{A} \) antennas communicates with \( K \) single-antenna users via a hybrid RF/RE EH-based UAV-RIS unit comprising \( \mathit{L} \) passive meta-surface elements. Over a finite horizon of \( T \) time slots, the objective is to maximize the EH efficiency while ensuring the minimum QoS requirements for all \( K \) users. This is achieved through efficient power allocation at the BS, optimal RIS phase shift control, and dynamic selection of reflective elements. More specifically, the optimization problem is stated as  
\normalsize{\allowdisplaybreaks
\begin{IEEEeqnarray}{lrC}
\IEEEyesnumber\label{eq:MOF_All} \IEEEyessubnumber*
\max_{\tau(t), \mathbf{P}_{U}, \bm{\beta}, \bm{\Theta}} \sum_{t=1}^{T} \varepsilon\varepsilon, \label{eq:MainOF} \\
\text{subject to}&\IEEEnonumber\\
R_k(t) \geq R_{\min}, \quad \forall k \in \mathbb{K}, t \in \mathbb{T}, \label{eq:C1} \\
0 \leq \tau (t) \leq 1, \quad \forall t \in \mathbb{T}, \label{eq:C2} \\
0 \leq P_{BS} \leq P_{BS}^{\max},  \label{eq:C3} \\
0 \leq p_k \leq P_U^{\max}, \quad \forall k \in \mathbb{K},  \label{eq:C4} \\
\beta_{i,j}^{k} \in \{0,1\}, ~ \forall i \in [1, \mathit{M}], j \in [1, \mathit{N}], k \in \mathbb{K}, \label{eq:C5} \\
\sum_{k \in \mathbb{K}} \beta_{i,j}^{k} \leq 1, \quad \forall i,j,  \label{eq:C6} \\
\theta_l \in [0,2\pi], \quad \forall l \in [1, \mathit{L}], \label{eq:C7} \\
|e^{j\theta_l}| = 1, \quad \forall l \in [1, \mathit{L}], \label{eq:C8} \\
\varepsilon^{r}(t) \geq \varepsilon^{c}(t), \quad \forall t \in \mathbb{T},  \label{eq:C9} \\
\varepsilon^{r}(t) \leq \varepsilon_{\max}, \quad \forall t \in \mathbb{T},  \label{eq:C10}
\end{IEEEeqnarray}}where the users' power allocation vector is denoted as \( \mathbf{P}_U = [p_1, p_2, \dots, p_K] \), and the RIS phase shift vector is given by \( \bm{\Theta} = [\theta_1, \theta_2, \dots, \theta_L] \) as each \( \theta_l \) represents the phase shift applied by the \( l \)-th RIS element. To dynamically manage resource allocation, the RIS selection matrix is defined as  
\begin{equation}
\bm{\beta} =
\begin{bmatrix}
\beta_{1,1}^1 & \dots & \beta_{1,\mathit{N}}^1 & \dots & \beta_{\mathit{M},\mathit{N}}^1 \\
\vdots & \ddots & \vdots & \ddots & \vdots \\
\beta_{1,1}^k & \dots & \beta_{1,\mathit{N}}^k & \dots & \beta_{\mathit{M},\mathit{N}}^k
\end{bmatrix}
\end{equation}
where each element \( \beta_{i,j}^{k} \) is a binary variable indicating whether RIS element \( (i,j) \) is assigned to user \( k \) for reflection.

The optimization problem is subject to several constraints. Constraint \eqref{eq:C1} ensures that each user $k$ maintains a minimum data rate \( R_{\min} \) to satisfy QoS requirements. Constraint \eqref{eq:C2} regulates the time slot division between EH and data transmission. The power constraints \eqref{eq:C3} and \eqref{eq:C4} limit the total transmit power at the BS and user terminals, ensuring EE. The RIS selection constraints \eqref{eq:C5} and \eqref{eq:C6} enforce binary selection of RIS elements for either reflection or EH, ensuring optimal element allocation. The RIS phase shift constraints \eqref{eq:C7} and \eqref{eq:C8} ensure that phase shifts remain within \( [0,2\pi] \) and adhere to passive beamforming properties. The causality constraint \eqref{eq:C9} guarantees that the UAV-RIS does not consume more energy than it has harvested at any given time slot. The battery overflow constraint \eqref{eq:C10} ensures that the stored energy does not exceed the UAV-RIS capacity \( \varepsilon_{\max} \).

The formulated problem \eqref{eq:MOF_All} is a MINLP problem with a non-convex objective function \eqref{eq:MainOF}, making it NP-hard. Consequently, it cannot be solved in a straightforward manner. So, we propose a DRL-based framework to deal with the problem detailed in the subsequent section.
\section{Joint Optimization DDPG-based Algorithm} \label{Sec_Proposed_Algorithm}
Conventional convex optimization techniques are often ineffective in ensuring long-term system performance, leading to suboptimal solutions. Moreover, the dynamic nature of wireless channel conditions requires adaptive optimization strategies that can efficiently respond to environmental variations. To address these challenges in UAV-assisted RIS systems, we propose a DRL-based approach, called the EE-DDPG algorithm, which is an enhanced variant of the TD3 algorithm \cite{fujimoto2018}. By leveraging the Markov Decision Process (MDP) framework, this method optimizes EH efficiency while maintaining communication quality. Specifically, it dynamically adjusts transmit power, RIS phase shifts, reflective element selection, and time slot allocation for EH and information transmission. The following sections define the MDP formulation within the DRL framework and introduce the proposed EE-DDPG algorithm.
\subsection{Markov Decision Process and Deep Reinforcement Learning: A DDPG-Based Framework}
To model the joint optimization problem, we formulate it as an MDP characterized by the tuple \( \langle A, S, P_t, R, \gamma \rangle \), where \( S \) represents the state space, \( A \) defines the available actions, \( P_t \) denotes the state transition probabilities, \( R \) is the instantaneous reward, and \( \gamma \) is the discount factor \cite{huang-2020}. The components are detailed as follows.
\subsubsection{State Space}
The state space integrates key system parameters, including the equivalent BS-to-UAV-RIS channel \( \mathbf{G}_1 \) and the UAV-RIS-to-user channel \( \mathbf{g}_{\phi,k} \in \mathbb{C}^{1\times\mathit{L}}, \forall k\in\mathcal{K} \). Additionally, it considers the positions of RIS elements \( \mathcal{P}^\phi_{i,j} \), user antennas \( \mathcal{P}^k \), and the harvested RE \( \varepsilon^{RE} \), enabling energy-aware decision-making and dynamic resource adaptation based on real-time energy availability. Besides, it includes the action of the previous time slot $a_{t-1}$. The system state is expressed as

{\footnotesize
\begin{align}
s_0 &= \big[ \Re\{\mathbf{G}_1\}, \Im\{\mathbf{G}_1\}, \Re \{\mathbf{g}_{\phi,k}\}, \Im \{ \mathbf{g}_{\phi,k}\}, \mathcal{P}^\phi_{i,j}, \mathcal{P}^k, \varepsilon^{RE}, a_{t-1} \big],
\end{align}}
where \( \Re \{\cdot\} \) and \( \Im \{\cdot\} \) denote the real and imaginary components.
\subsubsection{Action Space}
The action space consists of four primary control variables at each time slot \( t \): EH phase duration \( \tau(t) \in [0,1] \), power allocation for each user terminal \( p_k \), the reflective element selection variable \( \beta_{i,j}^{k} \), and the phase shift configuration of RIS elements \( \theta_l \). Among these, \( \tau(t) \), \( p_k \), and \( \theta_l \) are continuously adjustable, while \( \beta_{i,j}^{k} \) is binary. The complete action vector is
\begin{equation}
    a_t = \big[ \tau(t), p_k, \theta_l, \beta_{i,j}^{k} \big].
\end{equation}
\subsubsection{Reward Function}
The reward function \( r_t \) aims to maximize EH efficiency \( \varepsilon \varepsilon \) in \eqref{eq:MainOF} while ensuring compliance with system constraints in \eqref{eq:C1}--\eqref{eq:C10}. If any constraint is violated, the reward is set to zero, i.e., \( r_t=0 \). The state transition probability, \( P_t: S \times A \times S \to [0,1] \), defines the likelihood of transitioning from one state to another given an action. The discount factor, \( \gamma \in [0,1] \), determines the relative importance of future rewards versus immediate rewards.

At each decision step \( t \), the agent selects an action \( a_t = \mu^*(s_t) \) according to the optimal policy \( \mu^* \), based on the observed state \( s_t \). The agent obtains an immediate reward \( r_t = R(s_t, a_t) \) and transitions to the next state \( s_{t+1} \). The objective of reinforcement learning is to optimize \( \mu^* \) to maximize the expected cumulative reward
\begin{equation}
    \max_{\mu^*} J(\mu^*) := \mathbb{E} \left[ \sum_{t=0}^{T} \gamma^t r_t(s_t, \mu^*(s_t)) \right].
\end{equation}

The corresponding state-action value function, or Q-function, for policy \( \mu^* \) is
{\footnotesize
\begin{equation}
    Q^{\mu^*} (s_t, a_t) = \mathbb{E} \left[ \sum_{t=0}^{T} \gamma^t r_t \mid s_0 = s, a_0 = a, a_t \sim \mu^*(\cdot \mid s_t) \right].
\end{equation}
}
Since conventional Q-learning methods struggle with large continuous action spaces, DDPG is employed, leveraging an actor-critic architecture for efficient learning \cite{huang-2020}. The actor network learns a deterministic policy \( \mu(s \mid \vartheta^\mu) \), parameterized by \( \vartheta^\mu \), which maps states to actions. Meanwhile, the critic network evaluates the corresponding Q-values and provides feedback for policy improvement \cite{lillicrap-2016}. The critic network, \( Q(s, a \mid \vartheta^Q) \), parameterized by \( \vartheta^Q \), is trained using the Bellman equation. To ensure stable learning, target networks, delayed copies of the actor and critic networks, are introduced as \( \mu'(s \mid \vartheta^{\mu'}) \) and \( Q'(s, a \mid \vartheta^{Q'}) \). Exploration is facilitated using stochastic noise
\begin{equation}
    \mu'(s) = \mu (s \mid \vartheta^{\mu}) + \xi,
\end{equation}
where \( \xi \) represents a noise function tailored to the environment. The actor network updates its policy using the deterministic policy gradient
\begin{align}
    \nabla_{\vartheta^\mu} J \approx \frac{1}{N_b} \sum_{i} \big[ 
    & \nabla_a Q(s, a \mid \vartheta^Q) \mid_{s_i, a = \mu(s_i)} \nonumber \\
    & \nabla_{\vartheta^\mu} \mu(s \mid \vartheta^\mu) \mid_{s_i} \big],
\end{align}
where \( N_b \) denotes the mini-batch size sampled from the replay buffer \( \mathcal{D} \). The critic network updates using the loss function
\begin{equation}
    L(\vartheta^Q) = \frac{1}{N_b} \sum_{i=1}^{N_b} \left( y_i - Q(s_i, a_i \mid \vartheta^Q) \right)^2,
\end{equation}
where the target value is given by  
\begin{equation}
y_i = r(s_i, a_i) + \gamma Q'(s_{i+1}, \mu'(s_{i+1} \mid \vartheta^{\mu'}) \mid \vartheta^{Q'}).
\end{equation}

Finally, the target networks are updated using soft updates
\begin{align}
    \vartheta^{Q'} &\gets \varrho \vartheta^Q + (1 - \varrho) \vartheta^{Q'}, \nonumber\\
    \vartheta^{\mu'} &\gets \varrho \vartheta^\mu + (1 - \varrho) \vartheta^{\mu'},
\end{align}
where \( \varrho \ll 1 \) ensures smooth convergence and stable learning.
\subsection{The Proposed EE-DDPG Algorithm}
One of the primary limitations of the DDPG algorithm is its susceptibility to overestimation bias, which negatively impacts learning stability and policy performance. To mitigate this issue, the TD3 algorithm was introduced, incorporating a clipped double Q-learning mechanism \cite{fujimoto2018}. However, TD3 suffers from underestimation bias, leading to overly conservative policy updates that slow down convergence. To address both overestimation and underestimation biases, following \cite{peng-2023, jiang-2022}, we propose the EE-DDPG algorithm. It is an enhanced actor-critic framework that leverages dual actor networks and a softmax-weighted Q-value estimation mechanism.

The EE-DDPG framework consists of two independent actor networks, denoted as $ \mu_1(s \mid \vartheta^{\mu_1}) $ and $ \mu_2(s \mid \vartheta^{\mu_2}) $, each responsible for learning an optimal deterministic policy. These actors interact with two independent critic networks, $ Q_1(s, a \mid \vartheta^{Q_1}) $ and $ Q_2(s, a \mid \vartheta^{Q_2}) $, which estimate the action-value function using distinct learnable parameters. To ensure stable updates and prevent divergence, EE-DDPG maintains target versions of these networks, including a target actor $ \mu_1'(s \mid \vartheta^{\mu_1'}) $ and $ \mu_2'(s \mid \vartheta^{\mu_2'}) $, along with target critics $ Q_1'(s, a \mid \vartheta^{Q_1'}) $ and $ Q_2'(s, a \mid \vartheta^{Q_2'}) $. Unlike prior approaches that average the two main actors, EE-DDPG updates each target actor individually, ensuring stability while maintaining policy diversity.

To enhance exploration and prevent premature convergence, EE-DDPG modifies the target actor’s action by adding clipped Gaussian noise. At each time step, the next action is computed as $ a' = \mu_i'(s') + \epsilon $, where $i \in \{1,2\} $ denotes the selected target actor, and $ \epsilon \sim \mathcal{N}(0, \sigma) $. To ensure stability, the noise is clipped within a predefined range $ \epsilon = \text{clip}(\epsilon, -c, c) $, where $ c $ limits perturbation magnitude. This controlled exploration ensures policy robustness while mitigating excessive deviations. The noise-modified action is then used to compute the target Q-value, promoting reliable learning in dynamic environments.

Unlike TD3, which selects the minimum Q-value among critics to prevent overestimation, EE-DDPG refines the target Q-value estimation by applying a softmax-weighted expectation over the target critics. To achieve this, it first computes the conservative Q-value $Q_{\min}(s', a') = \min \left( Q_1'(s', a_1'), Q_2'(s', a_2') \right)$, then, it applies a softmax-weighted expectation over the target critics to balance exploration and exploitation as
\begin{equation}
    T_{\text{EE-DDPG}}(s') = \frac{\int_{a' \in A} \exp (\beta Q_{\min}(s', a')) Q_{\min}(s', a') da'}{\int_{a' \in A} \exp (\beta Q_{\min}(s', a')) da'},
    \label{eq:softmax}
\end{equation}
here, $ \beta $ is a temperature parameter that controls the influence of different Q-values in the weighting process. This formulation mitigates overestimation while ensuring that underestimation bias does not result in overly conservative policy updates. Once the softmax-weighted target Q-value is obtained, the target Q-value is computed using the Bellman equation
\begin{equation}
    y = r + \gamma T_{\text{EE-DDPG}}(s').
    \label{eq:Bellman}
\end{equation}
The critic networks are then trained by minimizing the Mean Squared Error (MSE) loss between the predicted Q-values and the target Q-value
\begin{equation}
    L(\vartheta^{Q_i}) = \mathbb{E} \left[ (y - Q_i(s, a \mid \vartheta^{Q_i}))^2 \right], \quad i \in \{1,2\}.
    \label{eq:critic_loss}
\end{equation}
This ensures that each critic is trained independently while leveraging the softmax-weighted target Q-value estimation to provide a robust and unbiased learning signal. In addition to critic updates, EE-DDPG incorporates a dual-actor learning mechanism, where two independent actors are maintained to enhance policy diversity. Rather than selecting an actor at random, EE-DDPG updates the actor corresponding to the critic that estimates the highest Q-value for the current state-action pair. This ensures that the policy associated with the more optimal value function is reinforced at each update step. At each training step, the action-value function is evaluated for both critics
\begin{equation}
    Q_i(s, a_i) = Q_i(s, \mu_i(s)), \forall{i,j}.
\end{equation}
The actor corresponding to the critic with the higher Q-value is then updated using the deterministic policy gradient
\begin{align}
    \vartheta^{\mu_{\text{chosen}}} &\leftarrow \vartheta^{\mu_{\text{chosen}}} + \alpha \mathbb{E} \bigg[
    \nabla_a Q(s, a \mid \vartheta^{Q_{\text{max}}}) \big|_{a=\mu_{\text{chosen}}(s)} \notag \\
    &\quad ~~\nabla_{\vartheta^{\mu_{\text{chosen}}}} \mu_{\text{chosen}}(s \mid \vartheta^{\mu_{\text{chosen}}}) 
    \bigg],
    \label{eq:actor_loss}
\end{align}
where the selected critic and actor pair is determined as
\begin{equation}
    (\mu_{\text{chosen}}, Q_{\text{max}}) =
    \begin{cases} 
        (\mu_1, Q_1), & \text{if } Q_1(s, a_1) \geq Q_2(s, a_2), \\
        (\mu_2, Q_2), & \text{otherwise}.
    \end{cases}
    \label{eq:chosen_actor}
\end{equation}
To stabilize training and avoid abrupt changes in learned policies, EE-DDPG employs a soft update mechanism for its target networks as follows
\begin{align}
    \vartheta^{Q_i'} &\gets \varrho \vartheta^{Q_i} + (1 - \varrho) \vartheta^{Q_i'}, \quad i \in \{1,2\}, \notag \\
    \vartheta^{\mu_i'} &\gets \varrho \vartheta^{\mu_i} + (1 - \varrho) \vartheta^{\mu_i'}, \quad i \in \{1,2\}.
    \label{eq:soft_upd}
\end{align}

This update ensures that each target actor evolves independently while maintaining training stability.
This final optimization mechanism ensures that the agent always selects the action that maximizes the expected return, corresponding to the highest EH efficiency \( \varepsilon\varepsilon \) in \eqref{eq:MainOF}. The EE-DDPG algorithm balances bias reduction, policy diversity, and stable learning while dynamically optimizing transmit power allocation \( p_k \), RIS phase shifts \( \theta_l \), reflective element selection \( \beta_{i,j}^{k} \), and EH time allocation \( \tau(t) \). By leveraging the state-action-reward formulation, the agent continuously refines its policy to adapt to dynamic channel conditions, ensuring sustained communication quality and efficient EH in UAV-assisted RIS systems. 

\subsubsection{UAV Trajectory Based on K-means Algorithms}
In dynamic environments with mobile users, the UAV-RIS must continuously adjust its position to ensure seamless communication. Operating at a fixed altitude, UAV-RIS optimizes its horizontal position within a Cartesian coordinate system to minimize path loss, which depends on the total Euclidean distance to all users. Our EE-DDPG algorithm adopts the density-aware deployment method \cite{peng-2023, lai-2019KM}, which positions the UAV-RIS to minimize the sum of squared geometric distances, enhancing connectivity and reducing path loss. The optimal horizontal position, \( \hat{\mathcal{P}}_\phi(t) \), is obtained by solving the following minimization problem
\begin{equation}
\min_{\hat{\mathcal{P}}_\phi(t)} \sum_{k\in \mathcal{K}} \|\hat{\mathcal{P}}_k(t) - \hat{\mathcal{P}}_\phi(t)\|^2,
\label{eq:k_mean}
\end{equation}
where \( \hat{\mathcal{P}}_k(t) \) denotes the horizontal position of user \( k \). The UAV-RIS position \( \hat{\mathcal{P}}_\phi(t) \) is computed using the K-means clustering algorithm detailed in \cite{dudik-2015}, which iteratively aligns with the user cluster centroid.
\begin{algorithm}[t]
\caption{Proposed EE-DDPG-Based Learning Algorithm}
\label{alg:EE-DDPG}
\begin{algorithmic}[1]
\STATE \textbf{Input:} $\mathbf{G}_1$, $\mathbf{g}_{\phi,k}$, $\forall{k\in\mathcal{K}}$, $\mathcal{P}^\phi_{i,j}$, $\forall{i,j}$,  $\mathcal{P}^k$, $\forall{k\in\mathcal{K}}$, $\varepsilon^{RE}$, experience replay size $\mathcal{R_D}$, episode count $N_E$, mini-batch size $N_b$, discount factor $\gamma$, soft update parameter $\varrho$.
\STATE \textbf{Output:} $a^* = \{\tau (t), \mathbf{P}_U, \bm{\beta}, \bm{\Theta}\}$ and the optimal EH efficiency $\varepsilon\varepsilon^*$ of the UAV-RIS system.
\STATE \textbf{Initialize:} Initialize two actor networks $\mu_1(s | \vartheta^{\mu_1})$ and $\mu_2(s | \vartheta^{\mu_2})$; two critic networks $Q_1(s, a | \vartheta^{Q_1})$ and $Q_2(s, a | \vartheta^{Q_2})$; target actor networks $\mu_1'(s | \vartheta^{\mu_1'}) = \mu_1$, $\mu_2'(s | \vartheta^{\mu_2'}) = \mu_2$; target critic networks $Q_1'(s, a | \vartheta^{Q_1'}) = Q_1$ and $Q_2'(s, a | \vartheta^{Q_2'}) = Q_2$; experience replay buffer $\mathcal{D}$ with size $\mathcal{R_D}$.
\FOR{each episode $N_E$}
    \STATE Obtain the latest CSI $\mathbf{G}_1$.
    \STATE Initialize the stochastic noise process $\xi$ for exploration.
    \STATE Acquire channel state vectors $\mathbf{g}_{\phi,k}$ for all $k \in \mathcal{K}$ in the $N_E$-th episode.
    \STATE Obtain the latest arrived RE $\varepsilon^{RE}$.
    \STATE Solve the K-means optimization problem in \eqref{eq:k_mean} to update the UAV-RIS position.
    \FOR{each time step $t$}
        \STATE Generate candidate actions $a_1 = \mu_1(s)$ and $a_2 = \mu_2(s)$.
        \STATE Select $a_{\text{chosen}}$ based on the critic with the highest Q-value:
        \STATE $(a_{\text{chosen}}, Q_{\text{max}}) = \begin{cases} a_1, Q_1, & \text{if } Q_1(s, a_1) \geq Q_2(s, a_2), \\ a_2, Q_2, & \text{otherwise}. \end{cases}$
        \STATE Apply exploration noise and execute action: $a_t = a_{\text{chosen}} + \epsilon$, where $\epsilon \sim \mathcal{N}(0, \sigma)$.
        \STATE Observe reward $r_t$, next state $s'$, and done flag $d_f$.
        \STATE Store the experience $(s, a_t, r_t, s', d_f)$ in $\mathcal{D}$.
        \STATE Sample a mini-batch of $N_b$ transitions from $\mathcal{D}$.
        \STATE Compute target action $a' = \mu_i'(s') + \text{clip}(\epsilon, -c, c)$ where $i \in \{1,2\}$.
        \STATE Compute conservative Q-value: $Q_{\min}(s', a') = \min(Q_1'(s', a_1'), Q_2'(s', a_2'))$.
        \STATE Compute softmax-weighted Q-value following \eqref{eq:softmax}.
        \STATE Compute target Q-value using \eqref{eq:Bellman}.
        \STATE Update each critic based on \eqref{eq:critic_loss}.
        \STATE Determine actor to update relying on \eqref{eq:chosen_actor}.
        \STATE Update the selected actor using the policy in \eqref{eq:actor_loss}.
        \STATE Perform soft updates for target networks as in \eqref{eq:soft_upd}.
    \ENDFOR
\ENDFOR
\end{algorithmic}
\end{algorithm}

Algorithm \ref{alg:EE-DDPG} details the proposed EE-DDPG scheme. The communication environment is represented by $\mathbf{G}_1$, $\mathbf{g}_{\phi,k}$, and $\mathcal{P}^\phi_{i,j}$ and $\mathcal{P}^k$ as the inputs in addition to the RE arrival $\varepsilon_{RE}$. Two actor networks, $\mu_1(s | \vartheta^{\mu_1})$ and $\mu_2(s | \vartheta^{\mu_2})$, and two critic networks, $Q_1(s, a | \vartheta^{Q_1})$ and $Q_2(s, a | \vartheta^{Q_2})$, are initialized along with their respective target networks. An experience replay buffer $\mathcal{D}$ of size $\mathcal{R_D}$ is used for learning.

At each episode, the agent updates the channel state $\mathbf{G}_1$, initializes a stochastic noise process $\mathcal{CN}$, and receives the current RE amount $\varepsilon^{RE}$. It then solves the K-means optimization problem to update the UAV-RIS position \( \mathcal{P}^\phi \) based on user distribution. The system state is then observed, and candidate actions $a_1 = \mu_1(s)$ and $a_2 = \mu_2(s)$ are generated. The action with the highest Q-value is selected using the corresponding critic, and exploration noise $\epsilon \sim \mathcal{N}(0, \sigma)$ is applied before execution. The agent receives an immediate reward $r_t$ and updates the replay buffer with the transition $(s, a_t, r_t, s', d_f)$. A mini-batch of $N_b$ samples is drawn from $\mathcal{D}$ for learning.

The target action is computed using the target actor network $\mu_i'(s')$, where $i \in \{1,2\}$, incorporating clipped noise for stability. The conservative Q-value is estimated using the minimum function before applying the softmax-weighted expectation over the target critics. The target Q-value is computed using the Bellman equation, and the critics are updated by minimizing the loss. The actor corresponding to the critic with the highest Q-value is updated using the policy gradient method.

Finally, the target networks are updated via soft updates with parameter \( \varrho \) to ensure smooth policy learning. The algorithm iterates until convergence, outputting the optimal control action \( a^* = \{\tau (t), \mathbf{P}_U, \bm{\beta}, \bm{\Theta}\} \) along with the optimal EH efficiency \( \varepsilon\varepsilon^* \) of the UAV-RIS system.
\begin{table}[!t]
\caption{Simulation Parameters} \label{tab:Simu}
\centering
\scalebox{0.9}{
\begin{tabular}{| c | c | c |}
\hline
Symbol & Description & Value\\
\hline
$A_{sim}$& Simulation area &$20$m × $20$m\\
\hline
$\mathit{L}$&  Number of RIS elements    &     $16$    \\ 
\hline
$C_X$ and $C_Y$ &LoS environmental constants & $9.61$ and $0.16$\\ 
\hline
$\upsilon$ &Reference path loss at {\footnotesize$d_{ref} = 1$} m & $-30$ dB\\ 
\hline
$\sigma_k^2$& User $k$ noise power      &     $-102$ dBm     \\ 
\hline
$\varphi$&   Non-LoS attenuation   &    $20$ dBm     \\  
\hline
$P_{BS}^{\max}$& BS max transmit power      &    $500$ W     \\ 
\hline
$\alpha$&  Path loss exponent for BS-RIS links    &     $3$    \\ 
\hline
$\nu$&  Path loss exponent for RIS-user links    &     $2.5$    \\ 
\hline
$R_{\text{min}} $ & Minimum user QoS requirements& $70$ Mbps\\
\hline 
\end{tabular}}
\end{table}
\section{Simulation Results} \label{Sec_Simu}
In this section, we evaluate the performance of the proposed EE-DDPG algorithm, considering the RF/RE EH capabilities of the RIS-enabled UAV network in both single-user (SU) and MU scenarios. For each scenario, we analyze the convergence performance of the EE-DDPG algorithm under the conventional TS EH strategy and the HERA strategy, both with and without RE (i.e., w/o RE), to demonstrate the role of RE in enhancing EH efficiency. Additionally, we compare the EH efficiency effectiveness of the proposed approach with three alternative models, namely the TD3, the DDPG, and the exhaustive search. This comprehensive analysis highlights the effectiveness and practicality of the proposed solution. Table \ref{tab:Simu} summarizes the simulation parameters used throughout this paper unless otherwise specified.

\subsection{Single-User Scenario Performance} \label{SubSec_Simu_SU}
\begin{figure}[!t]
\centering
\includegraphics[width=0.9\columnwidth]{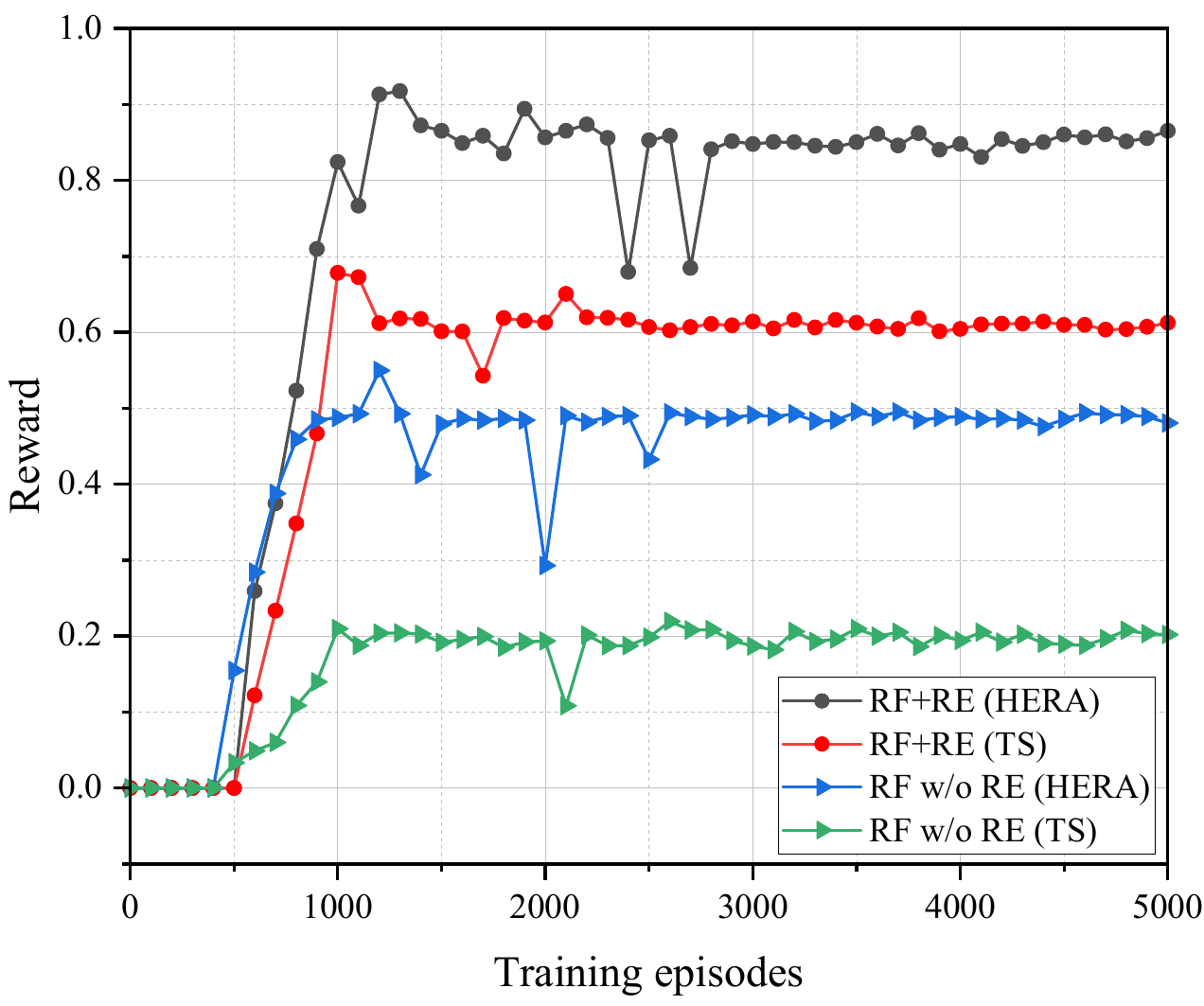}
\caption{Cumulative rewards per training episode for the SU scenario.}
\label{Fig_SU_Rewards}
\end{figure}
\begin{figure*}[!t]
\centering
\subfloat[]{\includegraphics[width=3.0in]{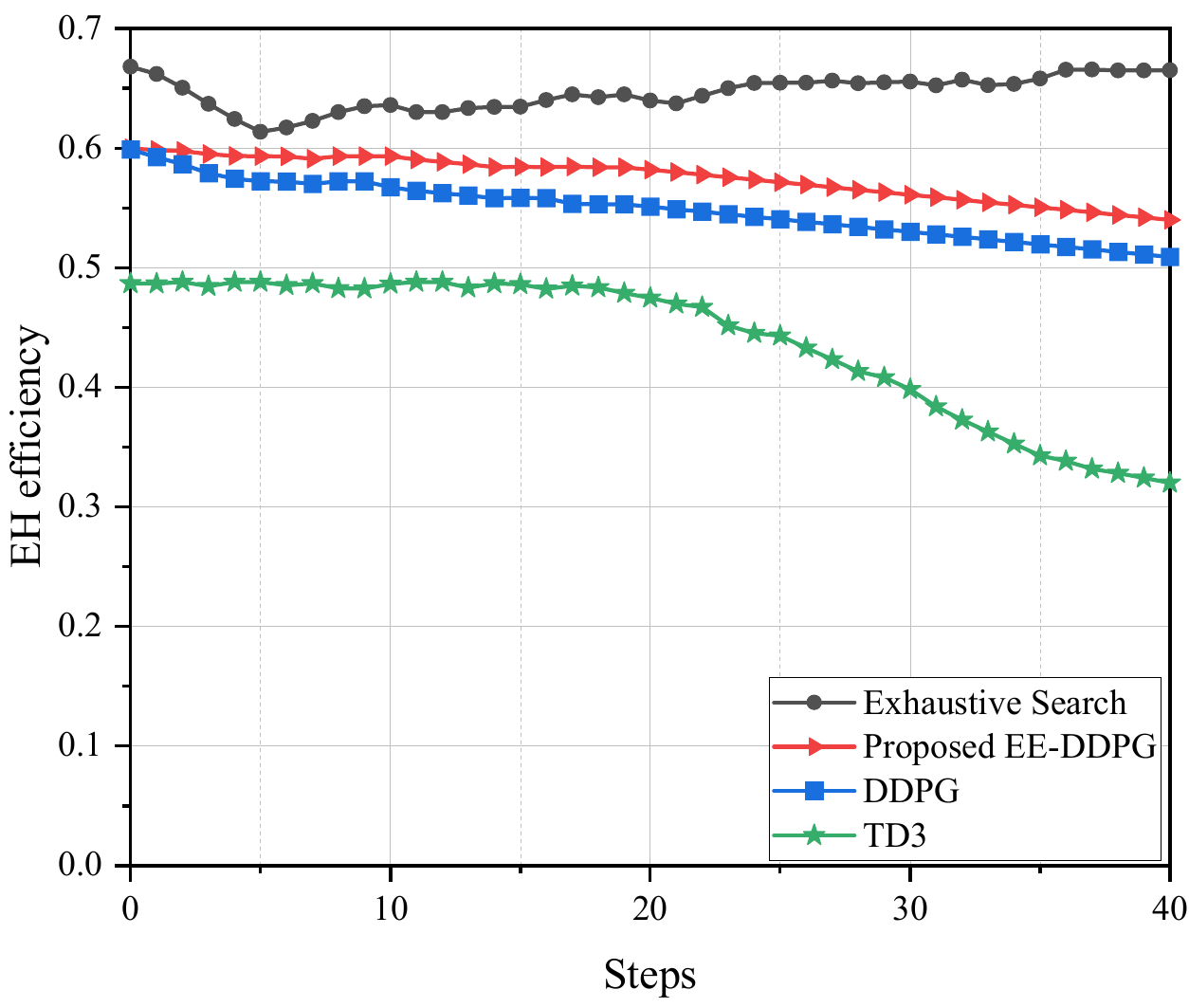}%
\label{Fig_SU_EH_vs_Steps_TS}}
\hfil
\subfloat[]{\includegraphics[width=3.0in]{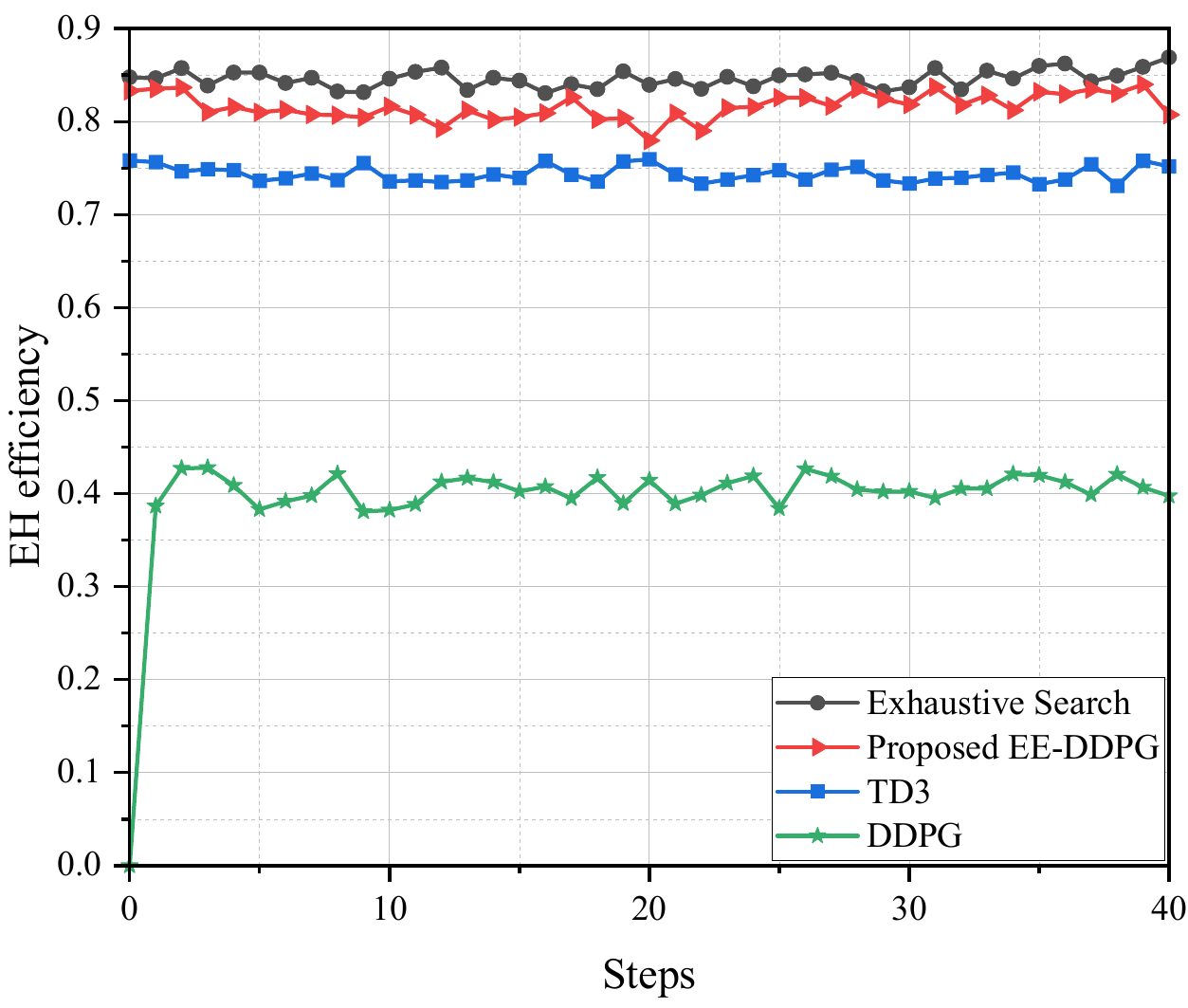}%
\label{Fig_SU_EH_vs_Steps_Hybrid}}
\caption{EH efficiency per testing step for the SU scenario. (a) TS strategy. (b) Proposed HERA strategy.}
\label{fig_SU_EH_TS&Hybrid}
\end{figure*}
Figure \ref{Fig_SU_Rewards} illustrates the convergence behavior of the proposed EE-DDPG algorithm in the SU scenario, considering nonlinear RF EH with and without RE for both the conventional TS and proposed HERA strategies. For the HERA strategy, the rewards initially increase, reaching approximately $0.8$ and $0.5$ per episode within the first $1,000$ episodes for RF EH with RE and without RE, respectively. However, the early training phase exhibits some instability. As training progresses, the rewards fluctuate between $0.67$ and $0.91$ for RF EH with RE and between $0.29$ and $0.54$ for RF EH without RE before gradually stabilizing after approximately $3,000$ episodes. Ultimately, the rewards converge around $0.86$ for RF EH with RE and $0.48$ for RF EH without RE, demonstrating the superior efficiency of the HERA strategy. In contrast, the conventional TS strategy exhibits a smoother and more consistent convergence pattern due to its lower problem complexity. The rewards increase steadily during the first $1,000$ episodes, reaching approximately $0.6$ and $0.2$ per episode for RF EH with RE and without RE, respectively. Unlike the HERA strategy, TS maintains stability throughout the training process. Beyond the initial $1,000$ episodes, the rewards converge smoothly, fluctuating within narrower ranges of $0.55$ to $0.66$ for RF EH with RE and $0.1$ to $0.2$ for RF EH without RE, achieving convergence in fewer than $2,500$ episodes. These results highlight the superior EH performance of the HERA strategy, particularly for RF EH with RE, while also emphasizing the faster convergence and stability of the TS strategy. Furthermore, the significant role of RE in enhancing EH efficiency is evident, as configurations incorporating RE consistently outperform those without RE. This demonstrates the importance of integrating RE sources in achieving higher rewards and improved overall performance.

The comparison of EH efficiency performance among various learning algorithms for the SU scenario is illustrated in Fig. \ref{fig_SU_EH_TS&Hybrid}, showing consistent results with the rewards in Fig. \ref{Fig_SU_Rewards}. The results are presented for the TS EH scheme in Fig. \ref{fig_SU_EH_TS&Hybrid}(a) and the proposed HERA scheme in Fig. \ref{fig_SU_EH_TS&Hybrid}(b) with RE. In both cases, the EE-DDPG algorithm performs slightly below the exhaustive search per step but achieves near-optimal performance in the HERA case. Although exhaustive search guarantees a globally optimal solution to the optimization problem, it is computationally intensive and lacks the adaptability required to handle dynamic environmental changes. As depicted in Fig. \ref{fig_SU_EH_TS&Hybrid}(a), the EE-DDPG algorithm outperforms both TD3 and DDPG in all steps in terms of EH efficiency under the TS strategy. However, DDPG surpasses TD3 due to the latter's underestimation issue, particularly when solving a less complex optimization problem. In contrast, as shown in Fig. \ref{fig_SU_EH_TS&Hybrid}(b), the TD3-optimized EH paradigm demonstrates more consistent performance than DDPG and achieves significant improvements in handling more complex scenarios.

Table \ref{tab:SU_EH} compares the average EH efficiency across different optimization schemes. The exhaustive search approach achieved the upper limits of $64.6\%$ and $84.6\%$ for the TS and HERA EH schemes, respectively. The proposed EE-DDPG method closely follows, harvesting $57.5\%$ and $81.5\%$ of the energy for the TS and HERA strategies, respectively. In addition, the TD3 approach achieved $43.9\%$ EH efficiency per step, while the DDPG-based system performed better under the TS scheme with $54.9\%$, benefiting from reduced susceptibility to underestimation issues. In contrast, the TD3 system achieved $74.3\%$ for the HERA scheme, significantly outperforming the DDPG method, which collected only $40.4\%$.

These results underscore the superior performance of the HERA strategy compared to the TS approach across all learning algorithms in SU scenario. Furthermore, the EE-DDPG model achieved the best performance among all learning algorithms for the TS EH scheme, demonstrating an optimal balance between effectiveness and practicality. While the exhaustive search approach achieved the highest efficiency, its computational complexity and lack of scalability render it impractical for real-world applications. Overall, the simulation results confirm the EE-DDPG model as the most effective and practical solution for the SU scenario.
\begin{table}[!t]
\caption{Average EH Efficiency for SU Scenario} \label{tab:SU_EH}
\centering
\scalebox{0.9}{
\begin{tabular}{|c|c|c|}
\hline
                   Algorithm & TS     & HERA   \\ \hline
Exhaustive search  & 64.6\% & 84.6\% \\ \hline
Proposed           & 57.5\% & 81.5\% \\ \hline
TD3 & 43.9\% & 74.3\% \\ \hline
DDPG               & 54.9\% & 40.4\% \\ \hline
\end{tabular}}
\end{table}
\subsection{Multi-User Scenario Performance} \label{SubSec_Simu_MU}
\begin{figure}[!t]
\centering
\includegraphics[width=0.9\columnwidth]{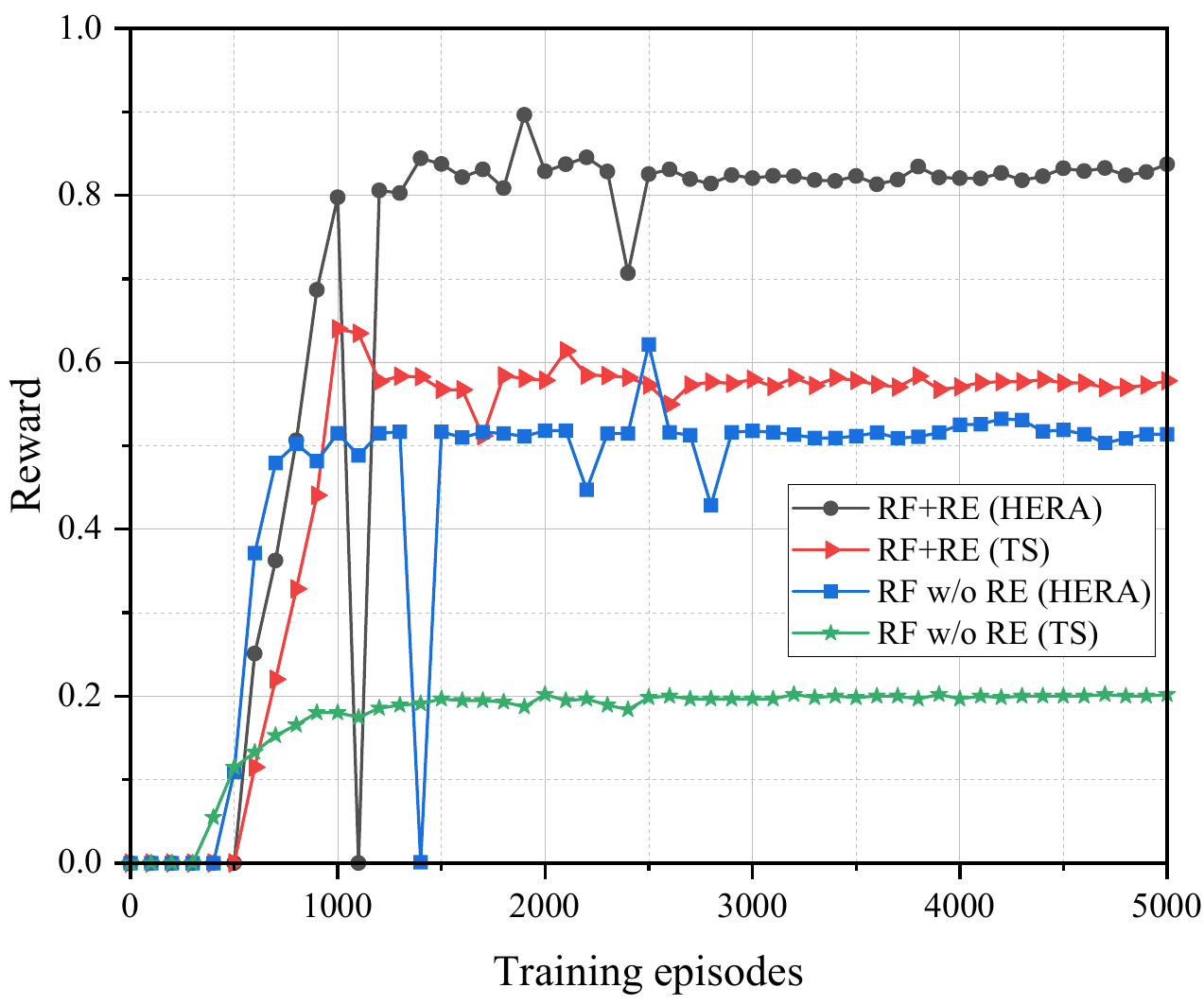}
\caption{Cumulative rewards per training episode for the MU scenario.}
\label{Fig_MU_Rewards}
\end{figure}

Figure \ref{Fig_MU_Rewards} depicts the convergence trend of the proposed EE-DDPG algorithm in the MU scenario for $K=3$, analyzing nonlinear RF EH with and without RE for both the proposed conventional TS and the HERA strategies.  
For the HERA strategy, the rewards increase initially, reaching approximately $0.78$ and $0.52$ per episode for RF EH with RE and without RE, respectively, within the first $1,000$ episodes. However, some instability is observed during the early training stages. As training continues, rewards fluctuate between $0.71$ and $0.9$ for RF EH with RE and between $0.43$ and $0.62$ for RF EH without RE before gradually stabilizing after around $3,000$ episodes. Finally, the rewards settle at approximately $0.82$ for RF EH with RE and $0.5$ for RF EH without RE, showcasing the HERA strategy's effectiveness in achieving high EH efficiency.  

In aligning with the SU scenario, the conventional TS strategy demonstrates a smoother and more consistent convergence pattern. The rewards steadily increase during the first $1,000$ episodes, reaching approximately $0.62$ and $0.18$ per episode for RF EH with RE and without RE, respectively. Unlike the HERA strategy, TS maintains stability throughout the training process. After the initial $1,000$ episodes, the rewards converge smoothly, fluctuating within narrower ranges of $0.5$ to $0.6$ for RF EH with RE and $0.19$ to $0.2$ for RF EH without RE. The TS strategy achieves convergence within around $2,500$ episodes. Again, these findings emphasize the HERA strategy's superior EH performance, particularly for RF EH with RE, while highlighting the TS strategy's faster convergence and greater stability. Similar to the SU scenario, the MU results highlight the significant role of RE in improving EH efficiency as our reward. RE enhances both the HERA and TS strategies boosting EH performance especially when satisfying the QoS of multiple users.

\begin{figure*}[!t]
\centering
\subfloat[]{\includegraphics[width=3.0in]{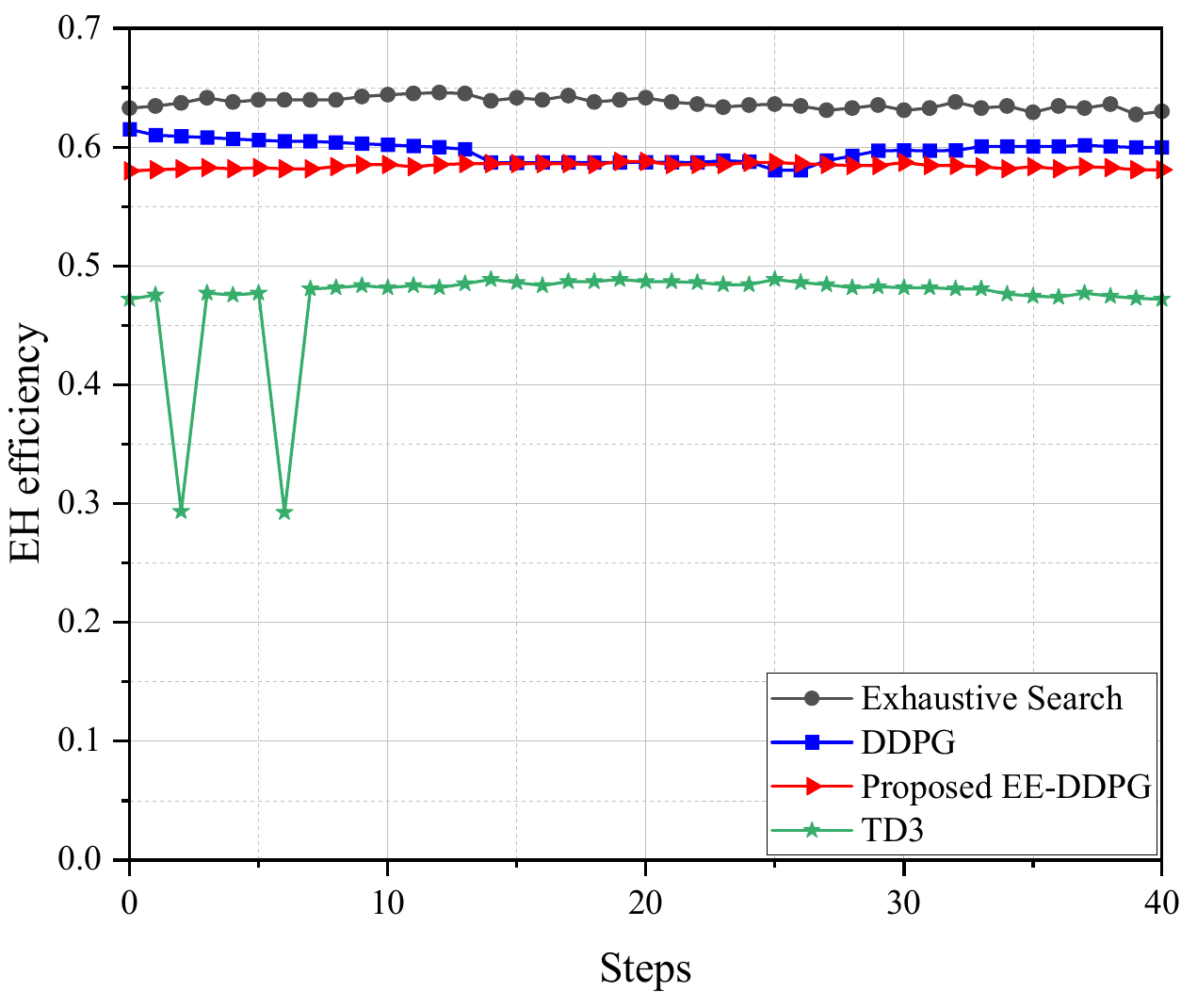}%
\label{Fig_MU_EH_vs_Steps_TS}}
\hfil
\subfloat[]{\includegraphics[width=3.0in]{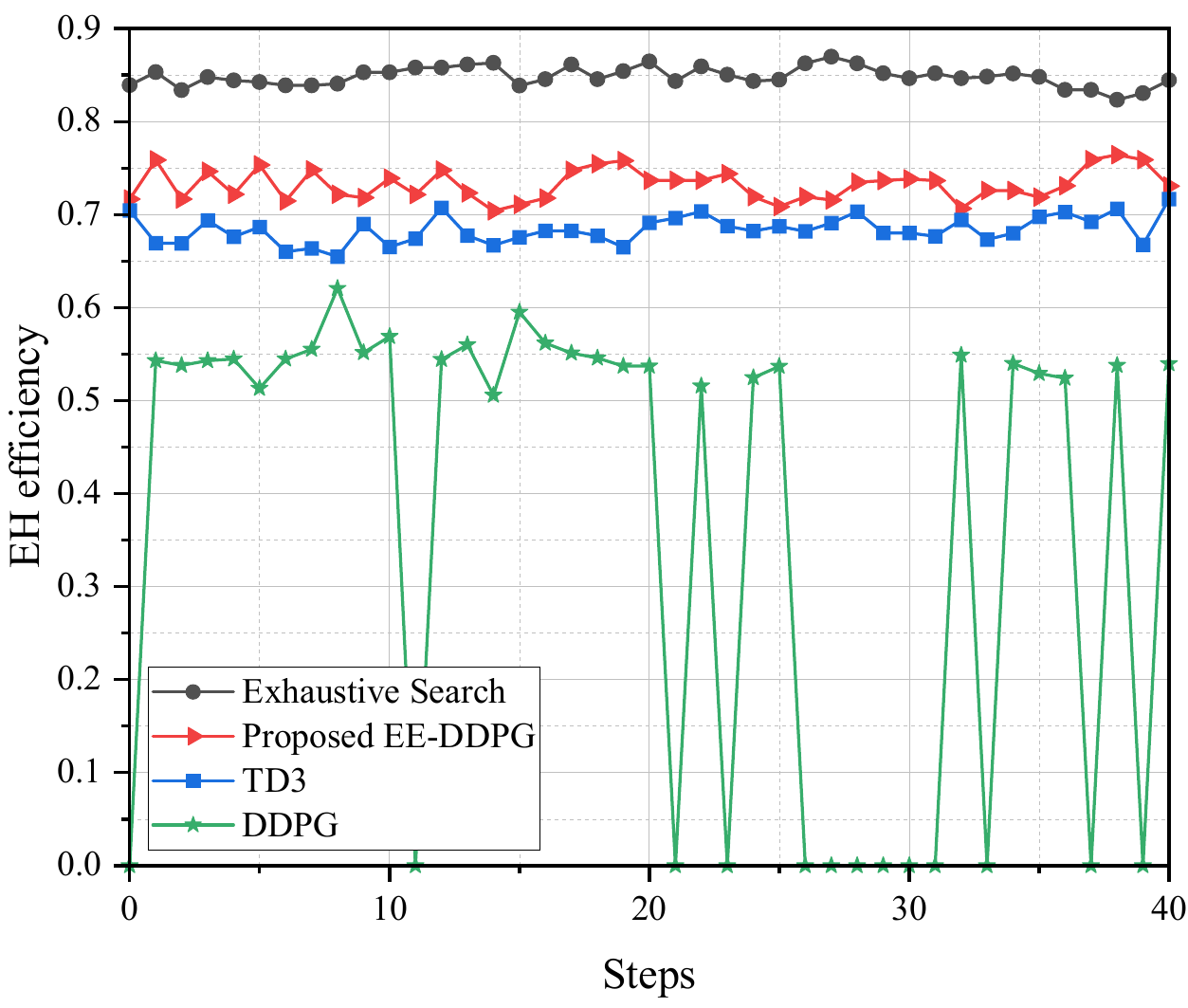}%
\label{Fig_SU_EH_vs_Steps_Hybrid}}
\caption{EH efficiency per testing step for the MU scenario ($K=3$). (a) TS strategy. (b) Proposed HERA strategy.}
\label{fig_MU_EH_TS&Hybrid}
\end{figure*}

The comparison of EH efficiency performance among various learning algorithms for the MU scenario is shown in Fig. \ref{fig_MU_EH_TS&Hybrid} corresponding to the rewards presented in Fig. \ref{Fig_MU_Rewards}. The results highlight the performance of the TS scheme in Fig. \ref{fig_MU_EH_TS&Hybrid}(a) and the proposed HERA scheme in Fig. \ref{fig_MU_EH_TS&Hybrid}(b) with RE. In both cases, the exhaustive search method demonstrates consistently higher EH efficiency than the other methods, as it explores all possible solutions to identify the optimal one, but in a time-consuming manner. As shown in Fig. \ref{fig_MU_EH_TS&Hybrid}(a), the proposed EE-DDPG and the DDPG models' performance is closely aligned, while the TD3 performs the worst. Notably, the simpler optimization problem in the TS scheme allows the DDPG algorithm to perform competitively outperforming the TD3 that suffers from slower learning due to its overestimation bias mitigation. On the other hand, as shown in Fig. \ref{fig_MU_EH_TS&Hybrid}(b), the performance of the proposed EE-DDPG closely follows that of the exhaustive search, with the TD3 coming next. However, the EH performance of the DDPG-based model lags behind. Here, the increased complexity of the HERA scheme highlights the superior adaptability and exploration capabilities of the TD3 algorithm, enabling it to outperform DDPG by mitigating overestimation bias and offering more stable performance.

Table \ref{tab:MU_EH} compares the average EH efficiency across different optimization schemes. The exhaustive search approach achieved the upper limits of $63.7\%$ and $84.8\%$ for the TS and HERA EH schemes, respectively. The proposed EE-DDPG method closely follows, harvesting $58.4\%$ and $73.2\%$ of the energy for the TS and HERA strategies, respectively. In addition, the TD3 approach achieved $47.2\%$ EH efficiency per step, while the DDPG-optimized approach performed better under the TS scheme with $68.4\%$, benefiting from reduced susceptibility to underestimation issues. In contrast, the TD3 system achieved $60.0\%$ for the HERA scheme, significantly outperforming the DDPG method, which collected only $37.2\%$. Aligning with the SU scenario, these results highlight the consistent superiority of the HERA strategy over the TS approach across all optimization methods. Additionally, the EE-DDPG model demonstrated robust performance, achieving an effective trade-off between learning efficiency and computational complexity, making it a practical choice for real-world MU scenarios.

\begin{figure}[!t]
\centering
\includegraphics[width=0.9\columnwidth]{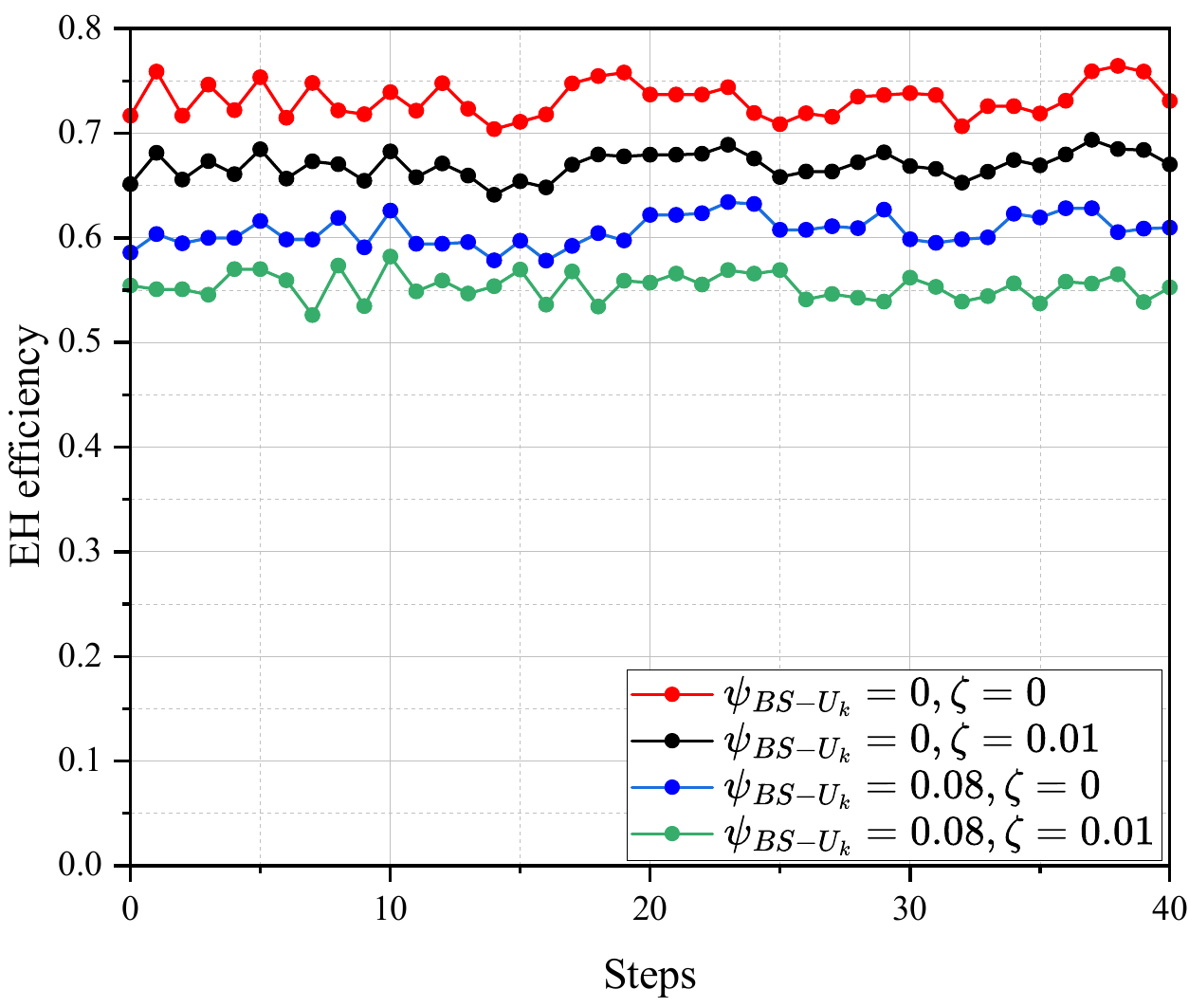}
\caption{Impact of channel and HIs on the EH efficiency of the proposed dual-DDPG model in the MU scenario ($K=3$) with HERA strategy.}
\label{Fig_MU_EH_Steps_Hybrid_Impairments}
\end{figure}

\begin{table}[!t]
\caption{Average EH Efficiency for MU ($K=3$) Scenario} \label{tab:MU_EH}
\centering
\scalebox{0.9}{
\begin{tabular}{|c|c|c|}
\hline
                   Algorithm & TS     & HERA   \\ \hline
Exhaustive search  & 63.7\% & 84.8\% \\ \hline
Proposed           & 58.4\% & 73.2\% \\ \hline
TD3 & 47.2\% & 68.4\% \\ \hline
DDPG               & 60.0\% & 37.2\% \\ \hline
\end{tabular}}
\end{table}
The impact of HIs and CSI imperfections on the proposed EE-DDPG algorithm considering the MU HERA strategy is depicted in Fig. \ref{Fig_MU_EH_Steps_Hybrid_Impairments}. The results demonstrate that the algorithm performs optimally under ideal conditions, achieving the highest EH when there are no impairments. However, the introduction of a CSI imperfection ($\zeta=0.01$) results in a marginal decline in efficiency, reflecting the algorithm's sensitivity to channel estimation errors. Furthermore, the presence of HIs ($\psi_{BS-U_k}=0.08$) leads to a more significant reduction in EH efficiency, indicating the challenges associated with managing and allocating resources in non-ideal hardware scenarios. The combination of both HIs and CSI imperfections ($\psi_{BS-U_k}=0.08,~\zeta=0.01$) increases the degradation, highlighting the compounded effects of these practical limitations. It is clear that these impairments negatively affect the users' QoS, necessitating additional energy to meet the required minimum data rate, which in turn impacts EH efficiency. 
\section{Conclusion} \label{Sec_Conclusion}
This paper investigated an energy-efficient UAV-assisted RIS communication system integrating hybrid RF and RE EH to extend UAV operational time. Through two phases, the proposed HERA strategy defined the RIS EH and element selection framework. However, the proposed EE-DDPG algorithm dynamically optimized RIS phase shifts, BS power allocation, TS parameter, and EH scheduling, accounting for RF EH non-linearity, CSI imperfections, and HIs. The EE-DDPG algorithm enhances learning stability and decision-making by incorporating action clipping and softmax-weighted Q-value estimation, which enables dynamic adaptation to user mobility. Simulation results validated the effectiveness of the proposed HERA strategy, demonstrating significant improvements in EH efficiency, thereby enhancing UAV sustainability. Additionally, the EE-DDPG model outperformed existing DRL algorithms while maintaining practical computational complexity. This work introduced an intelligent and robust algorithm capable of adapting to environmental dynamics and real-world imperfections. Future research directions include real-time UAV trajectory adaptation, cooperative multi-UAV networks, and further advancements in EH strategies to enhance the sustainability of UAV-RIS-assisted networks.
\bibliographystyle{IEEEtran}
\bibliography{References}

\begin{thebibliography}{10}
\providecommand{\url}[1]{#1}
\csname url@samestyle\endcsname
\providecommand{\newblock}{\relax}
\providecommand{\bibinfo}[2]{#2}
\providecommand{\BIBentrySTDinterwordspacing}{\spaceskip=0pt\relax}
\providecommand{\BIBentryALTinterwordstretchfactor}{4}
\providecommand{\BIBentryALTinterwordspacing}{\spaceskip=\fontdimen2\font plus
\BIBentryALTinterwordstretchfactor\fontdimen3\font minus \fontdimen4\font\relax}
\providecommand{\BIBforeignlanguage}[2]{{%
\expandafter\ifx\csname l@#1\endcsname\relax
\typeout{** WARNING: IEEEtran.bst: No hyphenation pattern has been}%
\typeout{** loaded for the language `#1'. Using the pattern for}%
\typeout{** the default language instead.}%
\else
\language=\csname l@#1\endcsname
\fi
#2}}
\providecommand{\BIBdecl}{\relax}
\BIBdecl

\bibitem{jiang-2021}
W.~Jiang, B.~Han, M.~A. Habibi, and H.~D. Schotten, ``{The road towards 6G: A comprehensive survey},'' \emph{IEEE Open J. Commun. Soc.}, vol.~2, pp. 334--366, Jan. 2021.

\bibitem{wang-2023}
C.-X. Wang, X.~You, X.~Gao, X.~Zhu, Z.~Li, C.~Zhang, H.~Wang, Y.~Huang, Y.~Chen, H.~Haas, J.~S. Thompson, E.~G. Larsson, M.~Di~Renzo, W.~Tong, P.~Zhu, X.~Shen, H.~V. Poor, and L.~Hanzo, ``{On the road to 6G: visions, requirements, key technologies, and testbeds},'' \emph{IEEE Commun. Surveys Tuts.}, vol.~25, no.~2, pp. 905--974, Jan. 2023.

\bibitem{Lin2025-Green}
N.~Lin, C.~Liu, T.~Wu, A.~Hawbani, L.~Zhao, S.~Wan, and M.~Guizani, ``Green communications: Ris-assisted fixed-wing uav coverage scheme based on deep reinforcement learning,'' \emph{IEEE Internet Things J.}, vol.~12, no.~4, pp. 4115--4127, Oct. 2025.

\bibitem{peng-2023}
H.~Peng and L.-C. Wang, ``{Energy harvesting reconfigurable intelligent surface for UAV based on robust deep reinforcement learning},'' \emph{IEEE Trans. Wirel. Commun.}, vol.~22, no.~10, pp. 6826--6838, Feb. 2023.

\bibitem{liu-2021S}
Y.~Liu, X.~Liu, X.~Mu, T.~Hou, J.~Xu, M.~Di~Renzo, and N.~Al-Dhahir, ``{Reconfigurable intelligent surfaces: principles and opportunities},'' \emph{IEEE Commun. Surveys Tuts.}, vol.~23, no.~3, pp. 1546--1577, Jan. 2021.

\bibitem{kurma-2023-DRL}
S.~Kurma, K.~Singh, P.~K. Sharma, and C.-P. Li, ``{DRL approach for spectral-energy trade-off in RIS-assisted full-duplex multi-user MIMO systems},'' \emph{Proc. IEEE Wireless Commun. Netw. Conf. (WCNC)}, Mar. 2023.

\bibitem{kumar-2024}
D.~Kumar, C.~K. Singh, O.~L.~A. López, V.~Bhatia, and M.~Latva-Aho, ``{Performance analysis of passive/active RIS aided wireless-powered IoT network with nonlinear energy harvesting},'' \emph{IEEE Trans. Wirel. Commun.}, p.~1, Dec. 2024.

\bibitem{lee2020deep}
G.~Lee, M.~Jung, A.~T.~Z. Kasgari, W.~Saad, and M.~Bennis, ``Deep reinforcement learning for energy-efficient networking with reconfigurable intelligent surfaces,'' in \emph{Proc. IEEE Int. Conf. Commun. (ICC)}.\hskip 1em plus 0.5em minus 0.4em\relax IEEE, Jun. 2020, pp. 1--6.

\bibitem{zhou-2023}
H.~Zhou, M.~Erol-Kantarci, Y.~Liu, and H.~V. Poor, ``{A survey on model-based, heuristic, and machine learning optimization approaches in RIS-aided wireless networks},'' \emph{IEEE Commun. Surveys Tuts.}, vol.~26, no.~2, pp. 781--823, Dec. 2023.

\bibitem{yu-2023}
Y.~Yu, X.~Liu, Z.~Liu, and T.~S. Durrani, ``{Joint trajectory and resource optimization for RIS assisted UAV cognitive radio},'' \emph{IEEE Trans. Veh. Technol.}, vol.~72, no.~10, pp. 13\,643--13\,648, Apr. 2023.

\bibitem{long2020reflections}
H.~Long, M.~Chen, Z.~Yang, B.~Wang, Z.~Li, X.~Yun, and M.~Shikh-Bahaei, ``Reflections in the sky: Joint trajectory and passive beamforming design for secure uav networks with reconfigurable intelligent surface,'' \emph{arXiv preprint arXiv:2005.10559}, 2020.

\bibitem{xu-2022Semi}
Y.~Xu, T.~Zhang, Y.~Zou, and Y.~Liu, ``{Reconfigurable Intelligence surface aided UAV-MEC systems with NOMA},'' \emph{IEEE commun. Lett.}, vol.~26, no.~9, pp. 2121--2125, Jun. 2022.

\bibitem{dhuheir2024multi}
M.~Dhuheir, A.~Erbad, A.~Al-Fuqaha, and M.~Guizani, ``Multi-uav multi-ris qos-aware aerial communication systems using drl and pso,'' in \emph{Proc. IEEE Int. Conf. Commun. (ICC)}.\hskip 1em plus 0.5em minus 0.4em\relax IEEE, Jun. 2024, pp. 654--659.

\bibitem{samir-2021}
M.~Samir, M.~Elhattab, C.~Assi, S.~Sharafeddine, and A.~Ghrayeb, ``{Optimizing age of information through aerial reconfigurable intelligent surfaces: a deep reinforcement learning approach},'' \emph{IEEE Trans. Veh. Technol.}, vol.~70, no.~4, pp. 3978--3983, Mar. 2021.

\bibitem{huang-2020}
C.~Huang, R.~Mo, and C.~Yuen, ``{Reconfigurable intelligent surface assisted multiuser MISO systems exploiting deep reinforcement learning},'' \emph{IEEE J. Sel. Areas Commun.}, vol.~38, no.~8, pp. 1839--1850, Jun. 2020.

\bibitem{cao-2023-DRL}
K.~Cao and Q.~Tang, ``{Energy efficiency maximization for RIS-Assisted MISO symbiotic radio systems based on deep reinforcement learning},'' \emph{IEEE commun. Lett.}, vol.~28, no.~1, pp. 88--92, Nov. 2023.

\bibitem{yang-2021}
Z.~Yang, M.~Chen, W.~Saad, W.~Xu, M.~Shikh-Bahaei, H.~V. Poor, and S.~Cui, ``{Energy-Efficient wireless communications with distributed reconfigurable intelligent surfaces},'' \emph{IEEE Trans. Wireless Commun.}, vol.~21, no.~1, pp. 665--679, Jul. 2021.

\bibitem{chen-2024}
X.~Chen, W.~Xu, Y.~Wang, D.~K. Yau, Y.~Li, and Z.~Cai, ``{Energy efficient optimization scheme for wireless powered active RIS in downlink system},'' \emph{IEEE Wirel. commun. Lett.}, p.~1, Jan. 2024.

\bibitem{mohammadi-2024}
M.~Mohammadi, H.~Q. Ngo, and M.~Matthaiou, ``{Phase-shift and transmit power optimization for RIS-Aided massive MIMO SWIPT IoT networks},'' \emph{IEEE Trans. Commun.}, p.~1, Jan. 2024.

\bibitem{ghose-2024}
S.~Ghose, A.~Kundu, D.~Mishra, S.~P. Maity, A.~Al-Nahari, and R.~Jäntti, ``{Energy efficient RIS-assisted wireless powered D2D communications in cognitive radio networks},'' \emph{IEEE Trans. Green Commun. Netw.}, p.~1, Jan. 2024.

\bibitem{sharma-2024}
N.~Sharma, S.~Gautam, S.~Chatzinotas, and B.~Ottersten, ``{Fractional programming based optimization techniques for RIS-assisted SWIPT-IoT system},'' \emph{IEEE commun. Lett.}, p.~1, Jan. 2024.

\bibitem{zhang-2023-RSMA}
R.~Zhang, K.~Xiong, Y.~Lu, P.~Fan, D.~W.~K. Ng, and K.~B. Letaief, ``{Energy efficiency maximization in RIS-assisted SWIPT networks with RSMA: a PPO-based approach},'' \emph{IEEE J. Sel. Areas Commun.}, vol.~41, no.~5, pp. 1413--1430, Jan. 2023.

\bibitem{wu-2020}
Q.~Wu and R.~Zhang, ``{Joint active and passive beamforming optimization for intelligent reflecting surface assisted SWIPT under QOS constraints},'' \emph{IEEE J. Sel. Areas Commun.}, vol.~38, no.~8, pp. 1735--1748, Jul. 2020.

\bibitem{hu-2024UAV}
J.~Hu, Y.~Ju, H.~Wang, L.~Liu, Q.~Pei, Y.~G. Shee, X.~Zhu, and C.~Wu, ``{UAV-RIS-aided energy-efficient and QOS-aware emergency communications based on DRL},'' \emph{Proc. IEEE 94th Veh. Technol. Conf. (VTC-Fall)}, pp. 1--5, Oct. 2024.

\bibitem{9771999}
H.~Peng, L.-C. Wang, G.~Ye~Li, and A.-H. Tsai, ``Long-lasting uav-aided ris communications based on swipt,'' in \emph{Proc. IEEE Wireless Commun. Netw. Conf. (WCNC)}, Apr. 2022, pp. 1844--1849.

\bibitem{fujimoto2018}
S.~Fujimoto, H.~Hoof, and D.~Meger, ``Addressing function approximation error in actor-critic methods,'' in \emph{Proc. 35th Int. Conf. Mach. Learn. (ICML)}.\hskip 1em plus 0.5em minus 0.4em\relax PMLR, Jul. 2018, pp. 1587--1596.

\bibitem{bisen-2021}
S.~Bisen, P.~Shaik, and V.~Bhatia, ``{On performance of energy harvested cooperative NOMA under imperfect CSI and imperfect SIC},'' \emph{IEEE Trans. Veh. Technol.}, vol.~70, no.~9, pp. 8993--9005, Sept. 2021.

\bibitem{sesia-2011}
S.~Sesia, I.~Toufik, and M.~Baker, \emph{{LTE - The UMTS Long Term Evolution: From Theory to Practice}}, Aug. 2011.

\bibitem{arzykulov-2019}
S.~Arzykulov, G.~Nauryzbayev, T.~A. Tsiftsis, B.~Maham, and M.~Abdallah, ``{On the outage of underlay CR-NOMA networks with detect-and-forward relaying},'' \emph{IEEE Trans. Cognit. Commun. Networking}, vol.~5, no. March, pp. 795--804, Sept. 2019.

\bibitem{salim-2020}
M.~M. Salim, D.~Wang, H.~A. E.~A. Elsayed, Y.~Liu, and M.~A. Elaziz, ``{Joint optimization of energy-harvesting-powered two-way relaying D2D Communication for IoT: A rate–energy efficiency tradeoff},'' \emph{IEEE Internet Things J.}, vol.~7, no.~12, pp. 11\,735--11\,752, Dec. 2020.

\bibitem{salim-2024}
M.~M. Salim, S.~I. Al-Dharrab, D.~B. Da~Costa, and A.~H. Muqaibel, ``{Rate-energy optimization for hybrid-powered full-duplex relays in cognitive C-NOMA with impairments},'' \emph{IEEE Open J. Commun. Soc.}, p.~1, Jan. 2024.

\bibitem{salim-2019Op}
M.~M. Salim, D.~Wang, Y.~Liu, H.~A. E.~A. Elsayed, and M.~A. Elaziz, ``{Optimal resource and power allocation with relay selection for RF/RE energy harvesting relay-aided D2D communication},'' \emph{IEEE Access}, vol.~7, pp. 89\,670--89\,686, Jan. 2019.

\bibitem{salim-2023AS}
M.~M. Salim, H.~A. Elsayed, and M.~S. Abdalzaher, ``{A survey on essential challenges in relay-aided D2D communication for next-generation cellular networks},'' \emph{J. Netw. Comput. Appl.}, vol. 216, p. 103657, Jul. 2023.

\bibitem{sharma-2022}
P.~K. Sharma, N.~Sharma, S.~Dhok, and A.~Singh, ``{RIS-assisted FD short packet communication with non-linear EH},'' \emph{IEEE commun. Lett.}, vol.~27, no.~2, pp. 522--526, Nov. 2022.

\bibitem{Semiha2023}
S.~Kosu, M.~Babaei, S.~Ã. Ata, L.~Durak-Ata, and H.~Yanikomeroglu, ``Linear/non-linear energy harvesting models via multi-antenna relay cooperation in v2v communications,'' \emph{IEEE trans. green commun. netw.}, vol.~7, no.~4, pp. 1725--1738, Jun.

\bibitem{peng-2021-Leo}
H.~Peng, A.-H. Tsai, L.-C. Wang, and Z.~Han, ``{LEOPARD: parallel optimal deep echo state network prediction improves service coverage for UAV-Assisted outdoor hotspots},'' \emph{IEEE Trans. Cogn. Commun. Netw.}, vol.~8, no.~1, pp. 282--295, Sept. 2021.

\bibitem{mei-2021}
H.~Mei, K.~Yang, J.~Shen, and Q.~Liu, ``{Joint trajectory-task-cache optimization with phase-shift design of RIS-assisted UAV for MEC},'' \emph{IEEE Wirel. commun. Lett.}, vol.~10, no.~7, pp. 1586--1590, Apr. 2021.

\bibitem{lei-2021}
M.~Lei, X.~Zhang, B.~Yu, S.~Fowler, and B.~Yu, ``{Throughput maximization for UAV-assisted wireless powered D2D communication networks with a hybrid time division duplex/frequency division duplex scheme},'' \emph{Wirel. Netw.}, vol.~27, no.~3, pp. 2147--2157, Feb. 2021.

\bibitem{al2014modeling}
A.~Al-Hourani, S.~Kandeepan, and A.~Jamalipour, ``Modeling air-to-ground path loss for low altitude platforms in urban environments,'' in \emph{Proc. IEEE Global Commun. Conf. (GLOBECOM)}.\hskip 1em plus 0.5em minus 0.4em\relax IEEE, Dec. 2014, pp. 2898--2904.

\bibitem{fadel2024irregular}
A.~S.~B. Fadel, T.~Y. Elganimi, K.~M. Rabie, G.~Nauryzbayev, and X.~Li, ``Irregular star-ris-aided wireless systems with a limited number of passive elements,'' in \emph{Proc. IEEE Int. Conf. Commun. Workshops (ICC Workshops)}.\hskip 1em plus 0.5em minus 0.4em\relax IEEE, Aug. 2024, pp. 1201--1206.

\bibitem{sohail2024irregular}
M.~U. Sohail, K.~M. Rabie, T.~Y. Elganimi, and G.~Nauryzbayev, ``Irregular ris-aided uav wireless systems in unstable environments,'' \emph{Authorea Preprints}, 2024.

\bibitem{salim-2022}
M.~M. Salim, H.~A. Elsayed, M.~A. Elaziz, M.~M. Fouda, and M.~S. Abdalzaher, ``{An optimal balanced energy harvesting algorithm for maximizing Two-Way relaying D2D communication data rate},'' \emph{IEEE Access}, vol.~10, pp. 114\,178--114\,191, Jan. 2022.

\bibitem{tang2020joint}
Y.~Tang, G.~Ma, H.~Xie, J.~Xu, and X.~Han, ``Joint transmit and reflective beamforming design for irs-assisted multiuser miso swipt systems,'' in \emph{Proc. IEEE Int. Conf. Commun. (ICC)}.\hskip 1em plus 0.5em minus 0.4em\relax IEEE, Jun. 2020, pp. 1--6.

\bibitem{lillicrap-2016}
\BIBentryALTinterwordspacing
T.~P. Lillicrap, J.~J. Hunt, A.~Pritzel, N.~Heess, T.~Erez, Y.~Tassa, D.~Silver, and D.~Wierstra, ``{Continuous control with deep reinforcement learning},'' \emph{arXiv (Cornell University)}, Jul. 2016. [Online]. Available: \url{https://arxiv.org/pdf/1509.02971.pdf}
\BIBentrySTDinterwordspacing

\bibitem{jiang-2022}
H.~Jiang, G.~Li, J.~Xie, and J.~Yang, ``{Action candidate driven clipped double Q-Learning for discrete and continuous action tasks},'' \emph{IEEE Trans. Neural Netw. Learn. Syst.}, vol.~35, no.~4, pp. 5269--5279, Sept. 2022.

\bibitem{lai-2019KM}
C.-C. Lai, C.-T. Chen, and L.-C. Wang, ``{On-demand density-aware UAV base station 3D placement for arbitrarily distributed users with guaranteed data rates},'' \emph{IEEE Wirel. commun. Lett.}, vol.~8, no.~3, pp. 913--916, Feb. 2019.

\bibitem{dudik-2015}
J.~M. Dudik, A.~Kurosu, J.~L. Coyle, and E.~Sejdić, ``{A comparative analysis of DBSCAN, K-means, and quadratic variation algorithms for automatic identification of swallows from swallowing accelerometry signals},'' \emph{Comput. Biol. Med.}, vol.~59, pp. 10--18, Apr. 2015.

\end{thebibliography}
\newpage
 




\vfill

\end{document}